\documentclass[pra,aps,showpacs,groupedaddress,superscriptaddress,twocolumn,toc=flat,biblatex,footinbib]{revtex4-1}
\usepackage{natbib}
\usepackage[table]{xcolor}
\usepackage[colorlinks=false]{hyperref}
\usepackage[utf8]{inputenc}
\usepackage{color}
\usepackage{bbm} 
\usepackage[english]{babel}
\usepackage{graphicx}
\usepackage{placeins}
\usepackage{multirow}
 \usepackage{amsmath} 
\usepackage{amssymb}
\usepackage{csquotes}

\usepackage{amsfonts,amsmath,amssymb,stmaryrd}

\usepackage{graphicx}
\usepackage{subfigure}  

\usepackage{epsfig}
\usepackage{mathrsfs}
\usepackage{verbatim}
\usepackage{centernot}
\usepackage{ulem}
\usepackage{longtable}
\usepackage{nicefrac}
\usepackage[framemethod=tikz]{mdframed}

\usepackage{array}

\usepackage{cancel,ifthen}
\newcommand{\cmnt}[2][NoInPuT]{\ifthenelse{\equal{#1}{NoInPuT}}{}{{\color{red}\sout{#1}}} {\color{blue} #2}}

\usepackage{bm}	
\renewcommand{\vec}[1]{\bm{#1}}

\LTcapwidth=\textwidth

\begin{document}

\normalem	

\title{Calculating spin-lattice interactions in ferro- and antiferromagnets: the role of symmetry, dimension and frustration}

\author{Hannah Lange}
\affiliation{Department of Chemistry/Phys. Chemistry, LMU Munich,
Butenandtstrasse 11, D-81377 Munich, Germany}

\author{Sergiy Mankovsky}
\affiliation{Department of Chemistry/Phys. Chemistry, LMU Munich,
Butenandtstrasse 11, D-81377 Munich, Germany}

\author{Svitlana Polesya}
\affiliation{Department of Chemistry/Phys. Chemistry, LMU Munich,
Butenandtstrasse 11, D-81377 Munich, Germany}
\author{Markus Weißenhofer}
\affiliation{Department of Physics, University of Konstanz, DE-78457 Konstanz, Germany}

\author{Ulrich Nowak}
\affiliation{Department of Physics, University of Konstanz, DE-78457 Konstanz, Germany}
\author{Hubert Ebert}
\affiliation{Department of Chemistry/Phys. Chemistry, LMU Munich,
Butenandtstrasse 11, D-81377 Munich, Germany}


\date{\today}

\begin{abstract}
Recently, the interplay between spin and lattice degrees of freedom has gained a lot of attention due to its importance for various fundamental phenomena as well as for spintronic and magnonic applications. Examples are ultrafast angular momentum transfer between the spin and lattice
subsystems during ultrafast demagnetization,  frustration driven by
structural distortions in transition metal oxides,
or in acoustically driven spin-wave resonances. In this
work, we provide a systematic analysis of spin-lattice interactions for
ferro- and antiferromagnetic materials and focus on the role of lattice symmetries and
dimensions, magnetic order, and the relevance of spin-lattice
interactions for angular momentum transfer as well as magnetic
frustration. For this purpose, we use a recently developed scheme which
allows an efficient calculation of spin-lattice interaction tensors from
first principles. In addition to that, we provide a more accurate and self
consistent scheme to calculate ab initio spin lattice interactions by
using embedded clusters which allows to benchmark the performance
of the scheme introduced previously. 
\end{abstract}

\maketitle

\section{Introduction}
The interplay of the magnetic and lattice degrees of freedom is crucial
for a number of phenomena observed for magnetic
materials.
Consequently, the investigation of spin-lattice interactions is of
great importance as it gives access to the understanding of a variety of
phenomena observed in magnetic materials, as for instance
frustration driven structural distortions in different transition metal
oxides  \cite{KG01, CPA+02, YRH+06} and dichalcogenides
\cite{Rasch2009,CRY+11}, or mutual modifications of 
the magnon and phonon spectra in the magnetically ordered state \cite{POL+16,OLN+16,KPLP19,VDF+20}.
Such a modification of the phonon spectra
can be seen for example by making use of Raman spectroscopy.
This was indeed demonstrated for multilayered CrI$_3$,
for which the modification of the Raman spectrum is associated with the
corresponding change of the phonon modes induced by
manipulating the interlayer alignment of magnetic moments in the
presence of a magnetic field \cite{MMU+20}.
In addition, a significant role of the the Dzyaloshinskii-Moriya interactions (DMI) in this material even without taking into account lattice vibrations has been demonstrated theoretically, with a strong impact of the DMI on the magnon spectrum \cite{Kvashnin2020,Solovyev}. Moreover, a crucial role of pronounced DMI-like spin-lattice interactions for the existence of topological magnons was also discussed on the basis of first-principles calculations in Ref. \cite{Sadhukhan2022}.
Interestingly, this phenomenon may be used for ultrafast optical control of
magnetism as discussed in the literature \cite{FSC+18, FTW+11}.
Furthermore, recent experimental and theoretical works show that
spin-lattice interactions play a crucial role for the angular momentum
transfer during ultrafast demagnetization \cite{Dornes, Tauchert2022}.
Apart from that, spin-lattice coupling has attracted increasingly attention during
last decade in view of its potential exploitation in spintronics and
magnonics, seen as a way to control magnetic
properties. This can be done for example by means of acoustic wave excitations, or via the application of external 
mechanical forces. In particular, spin-lattice interactions can be used for the control of the domain wall motion by optically
generated magnetoelastic waves \cite{OKB+15},
for spin current generation by surface acoustic waves in ferromagnetic
layers via magnon-phonon coupling (inverse Edelstein
effect \cite{XPA+18}), or for the control of the spin wave resonance frequency
by means of surface acoustic waves \cite{LLSL17}.

In this context the need for reliable schemes to investigate spin-lattice interactions as well as their dependence on the material 
under consideration emerges \cite{Garanin2015, Streib2019, Rueckriegel2020,Mentnik2019}. This field of research is only at
its beginning and to our knowledge no systematic investigation of the
role of spin-lattice coupling for a series of materials has been done
yet. A
promising way to gain insight in spin-lattice phenomena are atomistic
simulations that simultaneously model the time evolution of both spin and lattice degrees of freedom
\cite{Assmann2019, Ma2008, Perera2016,Strungaru2021}. This approach obviously requires a corresponding extension of the underlying model Hamiltonian to  account for the coupling between them \cite{Gar19,Hellsvik2019,Mankovsky2022}.
Accordingly, 
in order to
perform such simulations, besides spin-spin exchange coupling (SSC)
tensors also spin-lattice exchange coupling (SLC) tensors are needed as
an input. Recent works have provided first schemes to calculate these
tensors based on supercell and perturbative SLC approaches \cite{Hellsvik2019, Sadhukhan2022, Mankovsky2022}. 

In this work, an accurate and efficient method based
on embedded cluster calculations is presented and compared to the existing schemes. 
Furthermore, we calculate the SLC tensors for ferro- and antiferromagnetic materials
with different magnetic order, lattice structure and
dimensionality using the Korringa-Kohn-Rostoker (KKR) Green function method \cite{KKR} and systematically investigate the symmetry
of the SLC tensors w.r.t. the crystal symmetry as well as the physical
relevance of SLC to angular momentum transfer and magnetoelastic
transitions in frustrated antiferromagnets like metal-dichalcogenides
and -oxides.

To describe the coupling of spin and spatial degrees of freedom we adopt
the atomistic spin-lattice Hamiltonian as proposed by Hellsvik et
al. \cite{Hellsvik2019} and Mankovsky et al. \cite{Mankovsky2022}, i.e. 
\begin{eqnarray}
{\mathcal{ H}}_{sl} &=& - \sum_{i,j,\alpha,\beta} J_{ij}^{\alpha \beta}e_i^{\alpha}e_j^{\beta} 
  -  \sum_{i,j,\alpha,\beta} \sum_{k,\mu} {\mathcal{ J}}_{ij,k}^{\alpha \beta,\mu} e_i^{\alpha}e_j^{\beta}  u^{\mu}_k +\dots\,, 
\label{eq:Hamilt_extended_magneto-elastic}
\end{eqnarray}
%
with the spin-orientation vectors $\vec{e}_i$, atomic displacement
vectors $\vec{u}_k$, spin-spin coupling (SSC) tensor elements
$J_{ij}^{\alpha \beta}$ and spin-lattice coupling (SLC) tensor elements
${J}_{ij,k}^{\alpha \beta,\mu}=\frac{\partial J_{ij}^{\alpha
    \beta}}{\partial
  u_k^{\mu}}$. Eq. \eqref{eq:Hamilt_extended_magneto-elastic} can be
extended further to spin-lattice interactions of any order. Note that we
focus on magnetic interactions and hence omit pure 
lattice terms in Eq. \eqref{eq:Hamilt_extended_magneto-elastic} that
involve real space force constants.\\ 
\begin{figure}
    \centering
\includegraphics[width=0.38\textwidth]{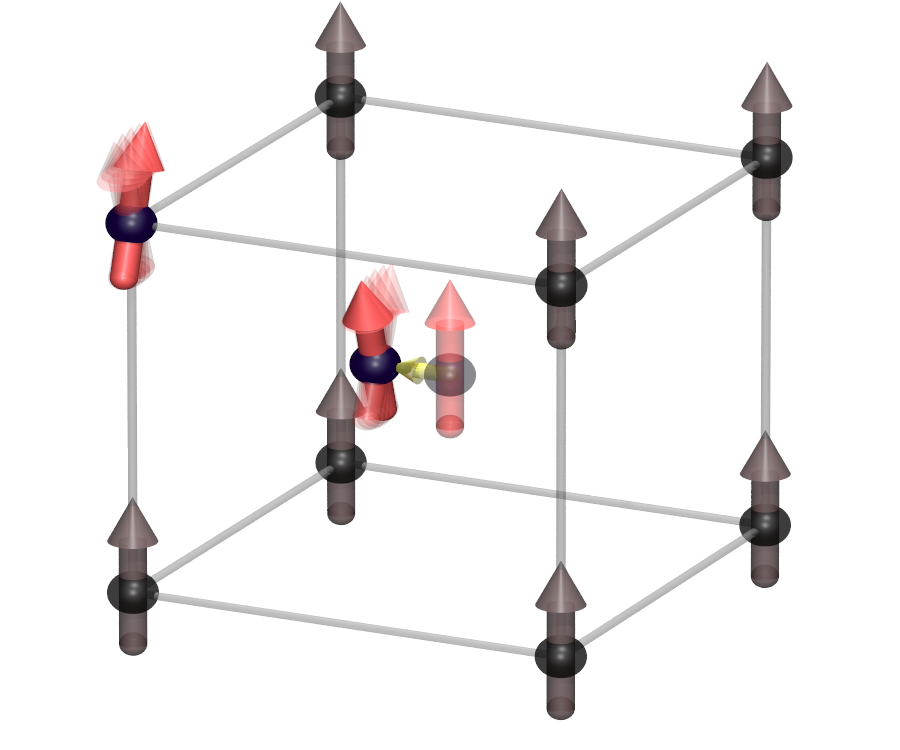}
    \caption{Geometry for the calculation of the spin-lattice exchange coupling using the embedded cluster method (here for $i=k$).}
    \label{fig:SLC}
\end{figure}

It should be mentioned that 
the mutual influence of spin and lattice dynamical properties
 have already been investigated previously
  using various approaches.
  A prominent example for this is the
  consideration of local spin-phonon interactions
  induced by a crystal field modulated due to a local
  lattice distortion \cite{ENR+76,AG71,TS72}.
  Another approach based on ab-initio  total energy calculations
  treats  the coupling parameters as
  fitting parameters \cite{Gar19, NWC+21}
  as done within the often used
  \textit{Conolly-Williams} approach \cite{CW1983}.
  In this case, ab-initio  electronic structure calculations
  have to be performed for a set of configurations large enough
  to fix all parameters for the chosen
  form of the Hamiltonian.
  Including the lattice in addition to the spin degrees of freedom
  obviously will make the fitting procedure much more demanding
  than considering a plain spin Hamiltonian. 

On the other hand,
the so-called Lichtenstein formula
\cite{Liechtenstein1987} is a well established approach for ab initio calculations of the isotropic spin-spin coupling parameters, with extensions to account for the full tensorial form of the interaction
parameters \cite{Udvardi2003, Ebert2009} and a multi-site expansion \cite{Mankovsky2021}. In contrast, the calculation of spin-lattice interaction parameters has received interest only recently. Refs. \cite{Hellsvik2019, Sadhukhan2022} have suggested to calculate the corresponding spin-lattice exchange coupling parameters from first principles by using supercells and calculating the SLC as the modification of the SSC when displacing an atom in each supercell. This method becomes accurate for sufficiently large supercells but is restricted to a small number of interacting atoms and relatively small supercells due to its high computational costs. Recently, Mankovsky et al. \cite{Mankovsky2022} have derived closed expressions to calculate the spin-lattice coupling tensors in a fully-relativistic way by treating the modifications in spin and lattice configurations on the same, perturbative level. It was shown that this method enables the calculation of fully relativistic SLC tensors which are in satisfying agreement with SLC tensors obtained by the supercell method.\\

The present paper is organized as follows:
In the first two sections, we briefly review the spin-lattice coupling methods presented in Ref. \cite{Mankovsky2022}, i.e. the supercell and perturbative method, and compare the results for bcc Fe to a new method based on embedded clusters, which enables a more efficient and accurate calculation of the SLC than the supercell method. Consequently we consider it as a more appropriate method to benchmark the perturbative SLC method presented in Ref. \cite{Mankovsky2022}, enabling a robust way of calculating spin-lattice interactions up to any order of displacements and spin tiltings. 
Comparing these results for all methods we find good agreement, which allows us to use the perturbative SLC method to systematically investigate spin-lattice coupling. In a first step, we consider the symmetry of the SLC tensors w.r.t. the crystal structure of the considered material and find that the symmetry of the lattice (in combination with spin-orbit coupling) gives rise to specific non-vanishing SLC contributions, e.g. Dzyaloshinskii–Moriya like interactions. In a second step, we calculate the SLC tensors for various materials with different lattice configurations and magnetic structures and consider the role of the dimensionality by calculating the SLC for free standing monolayers and 2D deposited magnetic films. Lastly, we investigate the SLC contribution to magnetoelastic transitions in frustrated antiferromagnets.

\section{Theoretical background}

The field of spin-lattice coupling is still at its beginning. The first
ones to calculate \textit{ab initio } spin-lattice interaction
parameters for the non-relativistic case were Hellsvik et
al.\ \cite{Hellsvik2019}, who suggested to calculate the elements of the
corresponding spin-lattice exchange coupling tensor from the
modification of the spin-spin exchange
coupling
$J_{ij}^{\alpha \beta}(\vec{u}_k)$ due to a displacement $u_k^\mu$. Focusing here on the spin-lattice
interaction term in the Hamiltonian, 
$\frac{\partial }{\partial u_k^\mu} J_{ij}^{\alpha \beta}(\vec{u}_k)\, u_k^\mu$,
linear with respect to displacement $u_k^\mu$,
the corresponding coupling parameters $J_{ij,k}^{\alpha\beta,\mu}$
can be delivered by calculating the  derivative of $J_{ij}^{\alpha
  \beta}(\vec{u}_k)$ numerically as follows 
\begin{align}
    J_{ij,k}^{\alpha\beta,\mu} =  \frac{\partial }{\partial u_k^\mu}
  J_{ij}^{\alpha \beta}(\vec{u}_k) \approx \frac{J_{ij}^{\alpha
  \beta}(u_k^\mu)-J_{ij}^{\alpha \beta}(0)}{u_k^\mu} 
    \label{eq:Cluster_SLC}
\end{align}
considering the displacement  $u_k^\mu$ in the limit of $u_k^\mu \to 0$.



As suggested by Hellsvik et al.\ \cite{Hellsvik2019},
the ordinary exchange coupling parameters $ J_{ij}^{\alpha \beta}(\vec{u}_k) $
can be calculated by making use of a
scheme introduced by  Liechtenstein and
coworkers leading to the so-called
Liechtenstein or LKAG formula  \cite{Liechtenstein1987}.
This approach that makes  use of the magnetic force theorem
implies the evaluation of the free energy
change due to a perturbation of the system, which can be written
within the multiple-scattering formalism \cite{KKR} as follows
\begin{eqnarray}
 \Delta {\cal F} &=&  -\frac{1}{\pi} \mbox{Im\, Tr\,} \int^{E_F}dE\,
                      \left(\mbox{ln}\, \underline{\underline{\tau}}(E) - \mbox{ln}\, \underline{\underline{\tau}}^{0}(E)\right) \; , 
\label{eq:Free_energy-2}
\end{eqnarray}
with the scattering path operator of the unperturbed reference system
\begin{eqnarray}
  \underline{\underline{\tau}}^{(0)}(E) &=&
          \Big[\underline{\underline{m}}^{(0)}(E) - \underline{\underline{G}}(E)\Big]^{-1}  \; ,
\label{eq:tau0}
\end{eqnarray}
and of the perturbed system
\begin{eqnarray}
  \underline{\underline{\tau}}(E) &=&
          \Big[\underline{\underline{m}}(E) - \underline{\underline{G}}(E)\Big]^{-1}  \; ,
\label{eq:tau}
\end{eqnarray}
with $\underline{\underline{G}}(E)$ the structural Green function and 
$\underline{\underline{m}}^{(0)}(E) =  [ \underline{\underline{t}}^{(0)}(E)  ]^{-1}$
the inverse of the  site-diagonal single-site
scattering matrix
$\underline{\underline{t}}^{(0)}(E)$, and  double underlines
indicating matrices with respect to site and spin-angular momentum indices. 

According to the LKAG scheme, the expression for the exchange
coupling parameter $J_{ij}$ is derived accounting for the perturbations
due to a spin tilting
$\delta \hat{e}_{i(j)}^\alpha$
on sites $i$ and $j$. As this perturbation leads to a
corresponding change of the inversed
single-site scattering matrix $\Delta^s_{\alpha}
\underline{m}_i = \underline{m}_i(\delta \hat{e}_i^\alpha) -
\underline{m}^0_i $, the change of the free energy in Eq.\
\eqref{eq:Free_energy-2} can be evaluated using the expression 
\begin{eqnarray}
  \mbox{ln} \,\underline{\underline{\tau}}
- \mbox{ln} \,\underline{\underline{\tau}}^0
  &=& - \ln \Big(1 +
       \underline{\underline{\tau}}\,[\Delta^s_{\alpha}
      {\underline{m}}_i + \Delta^s_{\beta}
      {\underline{m}}_j  +  ... ] \Big) \; .
\label{eq:tau2}
\end{eqnarray}
With the change of the inversed single-site scattering matrix 
represented in terms of so-called torque operator
$\underline{T}^{\mu}_i$ by the expression 
\begin{eqnarray}
  \Delta^s_{\mu} {\underline{m}}_i & = &  \delta \hat{e}^\alpha_i\,
       \underline{T}^{\alpha}_i \,,
                                         \label{eq:linear_distor_spin}
\end{eqnarray}
linear with respect to spin tilting, one obtains a direct access to the
exchange coupling parameters 
determined as the free energy derivative $\frac{\partial^2 {\cal F}}{\partial
  e^\alpha_i \,\partial e^\beta_j}$\cite{EM09a}:
\begin{eqnarray}  
  J_{ij}^{\alpha \beta}  &=&  -\frac{1}{\pi} \, \Im \, \mathrm{Tr} \,\int dE \, 
\underline{T}^{\alpha}_i \, \underline{\tau}^{ij}\,
\underline{T}^{\beta}_i  \, \underline{\tau}^{ji} \; .
                             \label{eq:LKAG}
\end{eqnarray}

To get access to the full exchange coupling tensor
Eq.\ (\ref{eq:LKAG}) has to be evaluated
within a fully relativistic framework \cite{Udvardi2003,Ebert2009}
with the  multiple-scattering
representation for the electronic Green function
$G(\vec{r},\vec{r}\,',E)$ in real space, given by the 
expression \cite{EBKM16} 
%
\begin{eqnarray}
G(\vec{r},\vec{r}\,',E) & = &
\sum_{\Lambda_1\Lambda_2} 
Z^{i}_{\Lambda_1}(\vec{r},E)
                              {\tau}^{i j}_{\Lambda_1\Lambda_2}(E)
Z^{j \times}_{\Lambda_2}(\vec{r}\,',E)
 \nonumber \\
 & & 
-  \sum_{\Lambda_1} \Big[ 
Z^{i}_{\Lambda_1}(\vec{r},E) J^{i \times}_{\Lambda_1}(\vec{r}\,',E)
\Theta(r'-r)  \nonumber 
\\
 & & \quad 
J^{i}_{\Lambda_1}(\vec{r},E) Z^{i \times}_{\Lambda_1}(\vec{r}\,',E) \Theta(r-r')
\Big] \delta_{ij} \, ,
\label{Eq_KKR-GF}
\end{eqnarray}
%
with the four-component wave functions $Z^{n}_{\Lambda}(\vec r,E)$
($J^{n}_{\Lambda}(\vec r,E)$) are regular (irregular) solutions to the
single-site Dirac equation \cite{MV79,ED11}.
The elements of the matrix $\underline{T}^{\mu}_i$ in
Eq.\ \eqref{eq:linear_distor_spin} are given by the expression
\begin{eqnarray}
 T^{\alpha}_{i,\Lambda\Lambda'} & = & \int_{\Omega_i} d^3r  \, Z^{i \times}_{\Lambda}(\vec{r},E)\, \Big[\beta \sigma_{\alpha} B_{xc}^i(\vec{r})\Big] \, Z^{
i}_{\Lambda'}(\vec{r},E)\,,  \label{Eq:ME-T-2}
\end{eqnarray}
with $B_{xc}^i(\vec{r})$ being the spin-dependent part of the
exchange-correlation potential, $\sigma_{\alpha}$ the Pauli matrix and
$\beta$ the standard Dirac matrix \cite{Ros61}.

\subsection{Super cell approach}

As demonstrated by  Hellsvik et al.\ \cite{Hellsvik2019}
and Mankovsky et al.\ \cite{Mankovsky2022},
Eq.\ (\ref{eq:LKAG})
can be used straightforwardly
to calculate  the exchange coupling parameter
$  J_{ij}^{\alpha \beta}(\vec{u}_k)$
in the presence of an atomic 
displacement on site $k$. Such calculations can be easily done also on the basis of the recently 
reported
approach based on Green’s functions constructed using Wannier 
functions
as a local basis set \cite{He2021}, 
that gives access
to an alternative way for the calculation of SLC parameters.
When performing these calculations using the multiple scattering formalism the 
scattering path operator $\underline{\underline{\tau}}(\vec{u}_k)$
is determined selfconsistently for
a supercell, big enough to minimize the impact on the exchange coupling tensor
$J_{ij}$ of the  displacement periodically repeated
in the  neighboring cells.
Note that selfconsistent calculations have a crucial impact on
the accuracy of the results, as in this case a relaxation of the charge
density around a displaced atom is taken into account. On the other
hand, an important disadvantage of supercell calculations is their
computational cost in the case of larger cells, or the other way around, they
lead to the increasing inaccuracy when the supercell size is too small.

\subsection{Embedded cluster approach \label{sec:EC}}

The disadvantages of the super cell approach -- high numerical costs
and a possible influence of neighboring super cells --
can be avoided by making use of the Dyson equation for
the Green function
\begin{eqnarray}
  {\cal G} = {\cal G}_0 + {\cal G}_0 \, \Delta {\cal V} \, {\cal G} \;
  \label{Dyson-Eq}
\end{eqnarray}
where ${\cal G}_0$ is the Green function of a suitable reference system,
while ${\cal G}$ accounts for the perturbation $ \Delta {\cal V}$.

To get access to the exchange parameter 
$  J_{ij}^{\alpha \beta}(\vec{u}_k)$
between site $i$ and $j$ for site $k$ displaced by $\vec{u}_k$
one considers an atomic cluster centered on site $k$
and big enough to include all sites  $i$ and $j$ of interest.
In a first step the Green function  ${\cal G}$
for this embedded cluster is calculated self-consistently
using Eq.\ (\ref{Dyson-Eq}) with $\Delta {\cal V}$ accounting
for the displacement of site  $k$ and its range given by the
size of the  embedded cluster.
Using the real space representation $G(\vec{r},\vec{r}\,',E)$
of the electronic Green function
given by Eq.\ (\ref{Eq_KKR-GF})
allows to replace the Dyson equation  (\ref{Dyson-Eq})
by the 
corresponding equivalent matrix equation for the scattering
path operators \cite{KKR}: 
%
\begin{equation}
   \underline{\underline{\tau}}_{k}(E) = [ (\underline{\underline{t}}_{k}(E))^{-1} 
- (\underline{\underline{t   }}_{0}(E))^{-1} - (\underline{\underline{\tau}}_{0}(E))
^{-1} ]^{-1}
\;.
\end{equation}
%
Here the second underline indicates matrices w.r.t.\ to the site indices numbering the sites within the cluster.
Accordingly,
the  scattering path operator matrices
$\underline{\underline{\tau}}_{0}(E)$ and $\underline{\underline{\tau}}_{k}(E)$
represent the unperturbed system
in the regime of the cluster
and the embedded cluster
with atom $k$ displaced by $\vec u_k$, respectively.
Finally, the single site matrices
$\underline{\underline{t}}_{0}(E)$ and $\underline{\underline{t}}_{k}(E)$
are site diagonal and represent the cluster atoms
in case of the  unperturbed system and the embedded cluster, respectively.

Having solved the embedding problem charge self-consistently
 the exchange coupling parameter
 $  J_{ij}^{\alpha \beta}(\vec{u}_k)$
 can be obtained from  Eq.\ (\ref{eq:LKAG})
 using the corresponding blocks
$ \underline{\tau}^{ij}(E)$ and $\underline{\tau}^{ji}(E)$ 
 of the super matrix
 $\underline{\underline{\tau}}_{k}(E)$.

\subsection{Perturbative approach}

Mankovsky et al. \cite{Mankovsky2022}
suggested a perturbative scheme to get direct acces
to the SLC parameter
$  J_{ij,k}^{\alpha\beta,\mu}$ without the numerical differentiation
indicated by  Eq.\ (\ref{eq:Cluster_SLC})
and to avoid this way high numerical effort and any spurious
inter-cell effects.
The central idea is to extend the scheme underlying
the Lichtenstein formula by accounting   simultaneously
for the impact of a distorted spin configuration $\{\delta
\hat{e}_i\}$ and of atomic displacements $\{\vec u_k\}$.
As a result, the change in the free energy
w.r.t.\ an unperturbed  reference state
is given in terms of the
scattering path operator the expression
\begin{eqnarray}
  \mbox{ln} \,\underline{\underline{\tau}}
- \mbox{ln} \,\underline{\underline{\tau}}^0
  &=& - \ln \Big(1 +
       \underline{\underline{\tau}}\,[\Delta^s_{\alpha}
      {\underline{m}}_i + \Delta^s_{\beta} {\underline{m}}_j + \Delta_{\mu}^u
                        {\underline{m}}_k +  ... ] \Big) \; ,
\end{eqnarray}
where $\Delta^u_{\mu} \underline{m}_k = \underline{m}_k(u_k^\mu) - \underline{m}^0_k $ is a change of the inverse single-site
scattering matrix due to atomic displacement on site $k$.
A linear approximation applied to $\Delta_{\mu}^u
\underline{m}_k$ representing it in terms of the so-called displacement
operator ${\cal U}_{k}^{\mu}$ \cite{Stefanou87,   Papanikolaou97} by the
expression 
\begin{eqnarray}
  \Delta^{u}_{\mu} \underline{m}_k & = & 
  u^\mu_k  \underline{\cal U}_{k}^{\mu}  \,
 \label{eq:linear_distor-disp}
\end{eqnarray}
allows us to work out explicit expression for the SLC parameters
$J^{\alpha\beta,\mu}_{ij,k}$ as
\begin{eqnarray}
  J^{\alpha\beta,\mu}_{ij,k} &= -\frac{\partial^3 {\cal F}}{\partial e^\alpha_i \,\partial e^\beta_j
\, \partial u^{\mu}_k} = -\frac{1}{2\pi} \mbox{Im\, Tr\,}
                                  \int^{E_F}dE \,\, \nonumber \\
                          \times &  \Big[   \underline{T}^{\alpha}_i \,\underline{\tau}_{ij}
                                  \underline{T}^{\beta}_j 
                                 \, \underline{\tau}_{jk}
                                \underline{\cal U}^{\mu}_k \,\underline{\tau}_{ki}
                             +  \underline{T}^{\alpha}_i
                                   \,\underline{\tau}_{ik}
       \underline{\cal U}^{\mu}_k \,\underline{\tau}_{kj}
                                  \underline{T}^{\beta}_j 
                                 \, \underline{\tau}_{ji}\Big] \;.
       \label{eq:Parametes_linear} 
\end{eqnarray}
The displacement operator in Eq.\ \eqref{eq:linear_distor-disp} is given
by the expression\cite{Mankovsky2022}
\begin{eqnarray}
 \underline{\cal U}_{k}^{\mu} & = & \bar{\underline{U}}(\hat{u}^{\,\mu}_k)\,\underline{m}_k+
\underline{m}_k\,\bar{\underline{U}}(-\hat{u}^{\,\mu}_k)\,,
\end{eqnarray}
where 
\begin{eqnarray*}
%
\bar{U}_{LL'}(\hat{u}_k)& = & \kappa \frac{4\pi}{3}i^{l+1-l'} 
\sum_{m=-1}^1 C_{LL'1m}  \,  Y_{1m}(\hat{u}_k) 
\end{eqnarray*}
and $\kappa = \sqrt{2mE/\hbar^2}$.
The prefactor $1/2$ occurs to avoid double counting of the identical
terms upon summations in Eq.\eqref{eq:Hamilt_extended_magneto-elastic}
over indices $i$ and $j$. In a similar way, higher order terms can be
expressed. The Fourier transforms of these parameters give access to the investigations of the impact of spin-lattice interactions on magnon and phonon modes (see Appendix \ref{appendix:FT}). Note however, that the SLC parameters given by Eq. \eqref{eq:Parametes_linear} do not account for the impact of screening of the atomic displacement due to the charge redistribution. To make sure that this contribution can be neglected with a reasonable accuracy of the results,
additional calculations discussed in Section \ref{sec:EC} have been performed to calculate the $J_{ij}$ parameters for the distorted lattice.  

Note that here we focus on the three-site exchange-like 
contributions to Eq.\ \eqref{eq:Hamilt_extended_magneto-elastic}, while a detailed discussion and benchmarking of 
the three-site approximation is presented in a complementary work \cite{Mankovsky2022_2}.  
This includes in particular technical details of higher order extensions to Eq.\ \eqref{eq:Hamilt_extended_magneto-elastic}.  Moreover, an expression for the SLC  parameters  that  
describe  a  modification  of the  mageto-cryslalline  anisotropy  due  
to  atomic displacements is presented and discussed together with 
corresponding numerical results.

\section{Results for the Embedded Cluster Approach}

\begin{figure}[t]
    \centering
\includegraphics[width=0.46\textwidth]{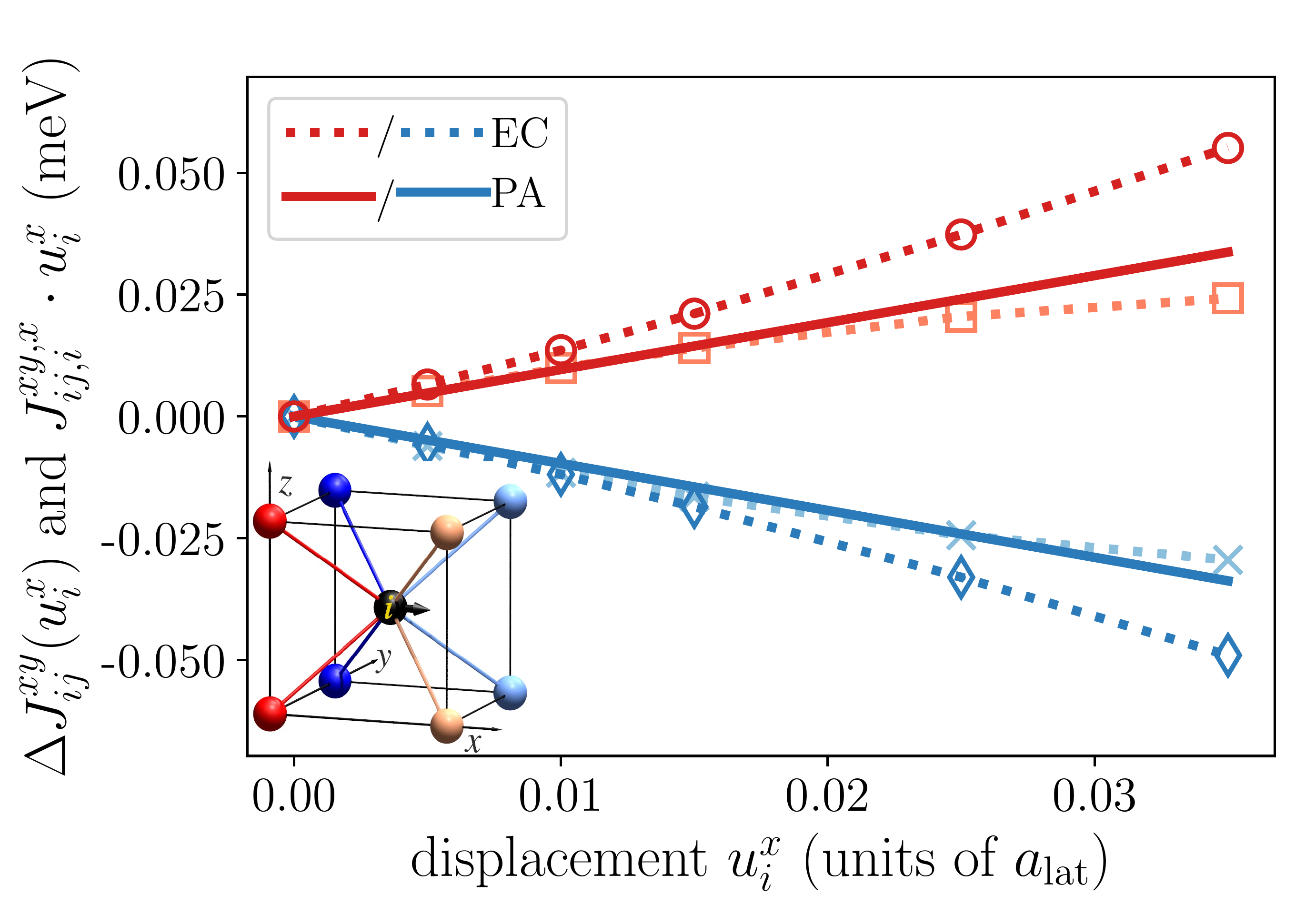}
    \caption{Comparison of the off-diagonal exchange coupling modification of nearest neighbors for embedded clusters  (EC) with 27 atoms $\Delta J_{ij}^{xy}(u_i^x)=J_{ij}^{xy}(u_i^x)-J_{ij}^{xy}(0)$ (dotted lines) and the perturbative approach (PA) $J_{ij,i}^{xy, x}\cdot u_i^x$ (solid lines), for a displacement of atom $i$ in $x$ direction for nearest neighbors $ij$ in bcc Fe. The color code for the atoms $j$ is visualized in the inset figure: Dark red circles denote neighbors with $\vec{r}_{ij}=a(-0.5, -0.5, \pm 0.5)$, light red squares $\vec{r}_{ij}=a(0.5, -0.5, \pm 0.5)$, dark blue diamonds $\vec{r}_{ij}=a(-0.5, 0.5, \pm 0.5)$ and light blue crosses $\vec{r}_{ij}=a(0.5, 0.5, \pm 0.5)$.}
    \label{fig:ClusterVsSLC_Plot2}
\end{figure}

In this section the properties of the SLC parameters obtained via the perturbative SLC method of Mankovsky et al. \cite{Mankovsky2022} and via the new method based on embedded cluster (EC) calculations are presented for bcc Fe. As for the supercell method the SLC $J_{ij,k}^{\alpha\beta,\mu}$ are obtained by the modification of the SSC $J_{ij}^{\alpha\beta}$ in the presence of a vanishingly small displacement $u_k^\mu$ \cite{Hellsvik2019} (see Fig. \ref{fig:SLC} and Eq. \eqref{eq:Cluster_SLC}) after a self-consistent (SCF) calculation of the potential for this distorted geometry has been done.

As mentioned above the cluster method has the advantage that it is accurate for finite cluster sizes, as long as the cluster is large enough to take into account the relaxation effects. This is already the case for relatively small system sizes, as can be seen in Tab. \ref{tab:SLC_1}, which shows that the results for diagonal and off-diagonal SLC parameters obtained from clusters consisting of $16$ and $51$ atoms are in very good agreement. In contrast, the top of table \ref{tab:SLC_1} shows that this is not the case for supercells consisting of $16$ and $54$ atoms. The supercell approach is in principle only accurate for infinite supercells since for finite sizes not only a single displaced atom is considered, but a periodic displacement for one atom in each supercell.

\begin{figure}
    \centering
    \includegraphics[width=0.46\textwidth]{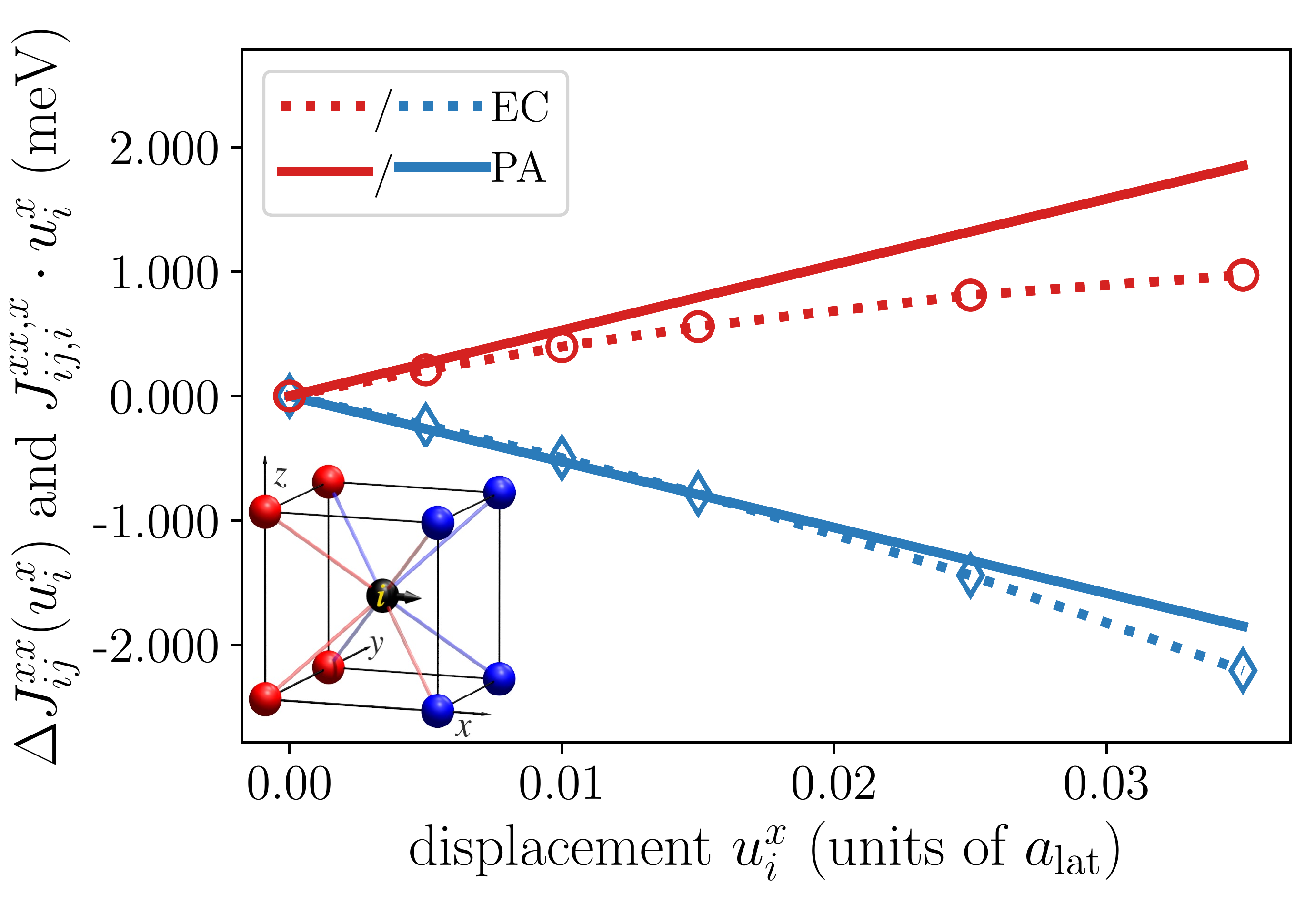}
    \caption{Comparison of the diagonal exchange coupling modification of nearest neighbors for embedded clusters  (EC) with 27 atoms $\Delta J_{ij}^{xx}(u_i^x)=J_{ij}^{xx}(u_i^x)-J_{ij}^{xx}(0)$ (dotted lines) and the perturbative approach (PA) $J_{ij,i}^{xx, x}\cdot u_i^x$ (solid lines), for a displacement of atom $i$ in $x$ direction for nearest neighbors $ij$ in bcc Fe. The color code of the atoms $j$ is visualized in the inset figure: Red circles correspond to neighbors with $\vec{r}_{ij}=a(-0.5, \pm(\mp)0.5, \pm 0.5)$ and blue diamonds to $\vec{r}_{ij}=a(0.5, \pm(\mp)0.5, \pm 0.5)$.}
    \label{fig:ClusterVsSLC_Plot1}
\end{figure}

Figures \ref{fig:ClusterVsSLC_Plot2} and \ref{fig:ClusterVsSLC_Plot1} as well as table \ref{tab:SLC_1} compare the results for nearest neighbors and $i=k$ obtained from supercell, cluster and perturbative SLC methods. More results for next-nearest neighbors and other SLC components are presented in Appendix \ref{appendix:NNN_EC}. In Fig. \ref{fig:ClusterVsSLC_Plot1} the modification of the cluster SSC occurring in Eq. \eqref{eq:Cluster_SLC}, i.e. $\Delta J_{ij}^{\alpha \beta}(u_k^\mu)=J_{ij}^{\alpha \beta}(u_k^\mu)-J_{ij}^{\alpha \beta}(0)$ is compared to the perturbative SLC result $J_{ij,k}^{\alpha \beta,\mu}\cdot u_k^\mu$ for $\alpha =\beta=\mu=x$. For small displacements, both results are in good agreement.
The average value of the diagonal components $J_{ij,i}^{\mathrm{diag-s},x}=\frac{1}{2}\left(J_{ij,i}^{xx,x}+J_{ij,i}^{yy,x} \right)$ is presented in Tab. \ref{tab:SLC_1}. Again, we find a good agreement between cluster and perturbative SLC results: The average absolute value for the large (small) cluster and nearest neighbors is $8.30\pm 0.22\,\mathrm{meV}/\mathrm{a.u.}$ ($8.28\pm 0.22\,\mathrm{meV}/\mathrm{a.u.}$), compared to $9.80\,\mathrm{meV}/\mathrm{a.u.}$ from the perturbative SLC method. The supercell results for the diagonal SLC are smaller with $7.02\pm 0.42\,\mathrm{meV}/\mathrm{a.u.}$ ($6.91\pm 0.23\,\mathrm{meV}/\mathrm{a.u.}$) for large (small) supercells. Similarly, the off-diagonal elements are in good agreement for all methods. 

The different values for different neighbors $j$ for both diagonal and off-diagonal components result primarily from higher order contributions taken into account by modifications of the electronic structure due to the displacements in clusters and supercells, but not in the perturbative approach. The results for the second neighbor shell are given in the Appendix \ref{appendix:NNN_EC}.  One can see a larger discrepancy  between the cluster and PA results, that can be associated with the important role 
of the screening effects in the exchange interactions at bigger distances indicating a long-range charge density redistribution around displaced atom. \\

To conclude, the embedded cluster method enables a very accurate calculation of the SLC parameters that can be used to benchmark the perturbative approach. We have shown that the results for bcc Fe agree well for both approaches.

\begin{table}
\centering
\begin{tabular}{c|cc|cc|c}
 $J_{ij,i}^{\mathrm{diag},x}$& \multicolumn{2}{c|}{supercells} & \multicolumn{2}{c|}{clusters}& \multicolumn{1}{c}{PA}\\
& $16$ atoms & $54$ atoms &$27$ atoms & $51$ atoms& \\\hline
$(-\frac{1}{2}, -\frac{1}{2}, \pm\frac{1}{2})$& 6.432&6.37 &7.80 &7.83 &9.80\\
$(-\frac{1}{2}, \frac{1}{2}, \pm\frac{1}{2})$& 6.432& 6.37&  7.82& 7.84&9.80\\
$(\frac{1}{2}, -\frac{1}{2}, \pm\frac{1}{2})$& -7.397&-7.67 &-8.76  &-8.77 &-9.80\\
$(\frac{1}{2}, \frac{1}{2}, \pm \frac{1}{2})$& -7.397& -7.67& -8.74 &-8.76 &-9.80\\
\end{tabular}
\vspace{0.5cm}\\

\begin{tabular}{c|cc|cc|c}
$J_{ij,i}^{\mathrm{off},x}$ & \multicolumn{2}{c|}{supercells} & \multicolumn{2}{c|}{clusters}& \multicolumn{1}{c}{PA}\\
& $16$ atoms & $54$ atoms &$27$ atoms & $51$ atoms& \\\hline
$(-\frac{1}{2}, -\frac{1}{2}, \pm\frac{1}{2})$& 0.23&0.21 &0.26 &0.25 &0.20\\
$(-\frac{1}{2}, \frac{1}{2}, \pm\frac{1}{2})$& -0.23& -0.21&  -0.22&-0.22&-0.20\\
$(\frac{1}{2}, -\frac{1}{2}, \pm \frac{1}{2})$& 0.21&0.19 &0.21  &0.21 &0.20\\
$(\frac{1}{2}, \frac{1}{2}, \pm\frac{1}{2})$& -0.21& -0.19& -0.24 &-0.24 &-0.20\\
\end{tabular}
\caption{Average absolute value of the diagonal SLC parameters $J_{ij,i}^{\mathrm{diag},x}=\frac{1}{2}\left(J_{ij,i}^{xx,x}+J_{ij,i}^{yy,x} \right)$ (top) and $J_{ij,i}^{\mathrm{off},x}=\frac{1}{2}\left(J_{ij,i}^{xy,x}+J_{ij,i}^{yx,x} \right)$ (bottom) in meV/a.u. for nearest neighbors, $i=k$ and a displacement in $x$ direction in bcc Fe obtained by the supercell method for supercells with $16$ and $54$ atoms, by the embedded cluster (EC) method for clusters with $27$ and $51$ atoms and for the closed SLC expressions. For the EC calculations $u_i^x = 0.005\,a_{\mathrm{lat}}$ was used. }
\label{tab:SLC_1}
\end{table}

\section{Analysis of Relativistic Spin-Lattice Exchange Coupling Tensors}
In this section, we will evaluate the connection of the spin-lattice exchange coupling tensors w.r.t. the crystal symmetry and dipole-dipole contributions based on analytical arguments, before discussing the numerical results in the next section.

\subsection{Symmetry of the SLC Parameters}

The qualitative features of the SLC parameters are determined by the symmetry of the system under investigation. Accordingly, we start our analysis of the spin-lattice exchange coupling tensors by linking their symmetry to the space group of the considered material. \\

The SLC tensor elements are connected by the relation
\begin{align}
    J_{ij,k}^{\alpha\beta, \mu} = \sum_{\alpha^{\prime}\beta^{\prime}, \mu^{\prime}} J_{i^{\prime}j^{\prime},k^{\prime}}^{\alpha^{\prime}\beta^{\prime},\mu^{\prime}}D(R)_{\alpha^{\prime}\alpha}D(R)_{\beta^{\prime}\beta}D(R)_{\mu^{\prime}\mu}
    \label{eq:Symmetry_Relation_Jijk}
\end{align}
for unitary and antiunitary symmetry transformations of the crystal, $u=\{R,\vec{p}\}$ and $a=T\{R,\vec{p}\}$ respectively, with $R$ denoting a rotation operation, $\vec{p}$ a primitive translation operation, $T$ the time inversion operation and $\underline{D}(R)$ the $3\times3$ matrix representation of $R$. The original and transformed site positions $\vec{r}_i$ and  $\vec{r}_{i^{\prime}}$ are related by $\vec{r}_{i^{\prime}}=R\vec{r}_i + \vec{p}$. Eq. \eqref{eq:Symmetry_Relation_Jijk} can explain many of the SLC properties observed in the previous section. As an example, we focus on next-nearest neighbors in bcc Fe, with the SLC presented in Appendix \ref{appendix:NNN_EC}  since they lie in $x,\,y,\,z$ directions in space and not in the diagonal directions like the nearest neighbors. For example, for $R$ being a 4-fold rotation around the $z$ axis the matrix representation is given by
\begin{align*}
    \underline{D}(R) = \begin{pmatrix}
0 & -1 & 0\\
1 & 0 & 0\\
0 & 0 & 1
\end{pmatrix}.
\end{align*}
Eq. \eqref{eq:Symmetry_Relation_Jijk} yields for $r_{ij}=(1,0,0)^T$ and $i=k$
\begin{align*}
    J_{ij,k}^{xy,x} =  J_{i^{\prime}j^{\prime},k^{\prime}}^{yx,y}\cdot 1\cdot (-1)\cdot 1 = -J_{i^{\prime}j^{\prime},k^{\prime}}^{yx,y}
\end{align*}
with $r_{i^{\prime}j^{\prime}}=(0,1,0)^T$. A similar argument holds for $J_{ij,k}^{yx,x}$ and $J_{ij,k}^{yx,y}$, as well as for the off-diagonal components $J_{ij,i}^{\mathrm{off},x}$ and $J_{ij,i}^{\mathrm{off},y}$. This is in agreement with table \ref{tab:SLC_4}. Furthermore, the relation can explain vanishing components in tables \ref{tab:SLC_4} in Appendix \ref{appendix:NNN_EC}, e.g. for $r_{ij}=(1,0,0)^T$ and $i=k$, for which a 2-fold rotation around the $z$ axis yields $J_{ij,k}^{xx,z}=J_{i^{\prime}j^{\prime},k^{\prime}}^{xx,z}$, but an inversion implies $J_{ij,k}^{xx,z}=-J_{i^{\prime}j^{\prime},k^{\prime}}^{xx,z}$ for $r_{i^{\prime}j^{\prime}}=(-1,0,0)^T$. Consequently, $J_{ij,k}^{xx,z}=J_{i^{\prime}j^{\prime},k^{\prime}}^{xx,z}=0$. 

In the following, we proof the symmetry relation \eqref{eq:Symmetry_Relation_Jijk}, starting with the expression for the SLC tensor as given in \eqref{eq:Parametes_linear}:
\begin{align}
    J_{ij,k}^{\alpha\beta, \mu} =  \frac{1}{2\pi}\mathrm{Im}\,\int^{E_\mathrm{F}}\mathrm{d}E\big[j_{ijk}^{\alpha\beta\mu,1}+j_{ijk}^{\alpha\beta\mu,2}\big]
\end{align}
with
\begin{align}  j_{ijk}^{\alpha\beta\mu,1}=\mathrm{Tr}\,\mathcal{T}_i^{\alpha}\tau_{ij}\mathcal{T}_j^{\beta}\tau_{jk}\mathcal{U}_k^{\mu}\tau_{ki}
\end{align}
and \begin{align}
    j_{ijk}^{\alpha\beta\mu,2}=\mathrm{Tr}\,\mathcal{T}_i^{\alpha}\tau_{ik}\mathcal{U}_k^{\mu}\tau_{kj}\mathcal{T}_j^{\beta}\tau_{ji}.
    \label{eq:jijk_1&2}
\end{align}
For an arbitary (unitary or antiunitary) symmetry operation $s$ the first part becomes
\begin{align*}
    j_{ijk}^{\alpha\beta\mu,1}=\mathrm{Tr}\,\mathcal{\tilde{T}}_i^{\alpha}\tilde{\tau}_{ij}\mathcal{\tilde{T}}_j^{\beta}\tilde{\tau}_{jk}\mathcal{\tilde{U}}_k^{\mu}\tilde{\tau}_{ki}
\end{align*}
with $\tilde{\mathcal{O}}_i^{\alpha}=s \mathcal{O}_i^{\alpha}s^{-1}$.
The scattering path operators $\tau$, torque operators $\mathcal{T}$ and displacement operators $\mathcal{U}$ behave under the different types of symmetry operations as follows:

For an unitary symmetry operation $u$ that transforms site $i(j)$ to $i^{\prime}(j^{\prime})$ the scattering path operator $\tau$ transformations as
$u\tau_{ij}u^{-1}=\tau_{i^{\prime}j^{\prime}}$ while for an antiunitary symmetry one has \cite{Huhne2002} $a\tau_{ij}a^{-1}=\tau_{j^{\prime}i^{\prime}}^{\dagger}$.
As pointed out in Ref. \cite{Seemann2015}, an arbitrary pseudovector transforms under a symmetry operation $s=\{R,\vec{p}\}$ or $s=T\{R,\vec{p}\}$ as 
    \begin{align*}
        s \vec{v}(\vec{r}) = \pm \mathrm{det}\left(\underline{D}(R) \right)\underline{D}(R)\vec{v}(s^{-1}\vec{r})
    \end{align*}
    while a vector transforms like
    \begin{align*}
        s \vec{v}(\vec{r}) = \pm \underline{D}(R)\vec{v}(s^{-1}\vec{r})\,.
    \end{align*}
    In these expressions the sign $\pm$ is determined by the behavior of $\vec{v}$ under time reversal: the positive sign applies for (polar) vectors and the negative sign for (axial) pseudovectors \cite{Seemann2015}. As the torque operator $\mathcal{\tilde{T}}$ behaves as a pseudovector the sign $\pm$ and $\mathrm{det}\left(\underline{D}(R) \right)=\pm 1$ from the first expression occur twice in Eq. \eqref{eq:jijk_1&2}, and hence cancel each other. Consequently, one finds
    \begin{align}
        \mathcal{\tilde{T}}_i^{\alpha}\dots \mathcal{\tilde{T}}_j^{\beta}=\sum_{\alpha^{\prime}\beta^{\prime}}\mathcal{T}_{i^{\prime}}^{\alpha}\dots\mathcal{T}_{j^{\prime}}^{\beta}D(R)_{\alpha^{\prime}\alpha}D(R)_{\beta^{\prime}\beta}.
    \end{align}
    For the displacement operator, which is not affected by time reversal, we have a positive sign and hence
    \begin{align}
        \mathcal{\tilde{U}}_k^{\mu}=\sum_{\mu^{\prime}}\mathcal{U}_{k^{\prime}}^{\mu^{\prime}}D(R)_{\mu^{\prime}\mu}.
    \end{align}

For unitary operations this yields
\begin{align*}
    j_{ijk}^{\alpha\beta\mu,1}= \sum_{\alpha^{\prime}\beta^{\prime}, \mu^{\prime}} j_{i^{\prime}j^{\prime}k^{\prime}}^{\alpha^{\prime}\beta^{\prime}\mu^{\prime},1}D(R)_{\alpha^{\prime}\alpha}D(R)_{\beta^{\prime}\beta}D(R)_{\mu^{\prime}\mu}.
\end{align*}
For antiunitary operations we find
\begin{align*}
    j_{ijk}^{\alpha\beta\mu,1}
    = \sum_{\alpha^{\prime}\beta^{\prime}, \mu^{\prime}} j_{i^{\prime}j^{\prime}k^{\prime}}^{\alpha^{\prime}\beta^{\prime}\mu^{\prime},2}D(R)_{\alpha^{\prime}\alpha}D(R)_{\beta^{\prime}\beta}D(R)_{\mu^{\prime}\mu}
\end{align*}
and
\begin{align*}
    j_{ijk}^{\alpha\beta\mu,2}= \sum_{\alpha^{\prime}\beta^{\prime}, \mu^{\prime}} j_{i^{\prime}j^{\prime}k^{\prime}}^{\alpha^{\prime}\beta^{\prime}\mu^{\prime},1}D(R)_{\alpha^{\prime}\alpha}D(R)_{\beta^{\prime}\beta}D(R)_{\mu^{\prime}\mu}\,.
\end{align*}
Hence, the same relation for unitary as well as antiunitary symmetry operations holds.

\subsection{Dipole-Dipole Contribution to SLC}
\begin{table}
\centering
\begin{tabular}{c|c|cccc}
& & Fe ($r_{ij}=0.87$) &Fe ($r_{ij}=1$) & MnGe &Au on Fe \\\hline
\multirow{8}{*}{SSC} 
& $J_{ij}^{\mathrm{diag-s}}$&18.051& 10.090& 18.187&21.414\\
&$J_{ij}^{\mathrm{diag-a}}$&0.0&0.015&0.009&0.0091\\ &$J_{ij}^{\mathrm{off-s},x}$&0.013&0.0&0.012&0.079\\
&$\vert \vec{D}_{ij}^{x}\vert$& 0.0&0.0&0.151&0.272\\\cline{2-6}
&$J_{ij,\mathrm{dip}}^{\mathrm{diag-s}}$&0.0&0.013&0.004&0.006\\
& $J_{ij,\mathrm{dip}}^{\mathrm{diag-a}}$&0.00&0.039&0.006&0.008\\
&$J_{ij,\mathrm{dip}}^{\mathrm{off-s},x}$& 0.040&0.0&0.008&0.015 \\
&$\vert \vec{D}_{ij,\mathrm{dip}}^{x}\vert$&0.0&0.0&0.0&0.0\\\hline
\multirow{8}{*}{SLC} 
& $J_{ij,i}^{\mathrm{diag-s}}$&9.792& 1.858& 9.792&7.693\\
&$J_{ij,i}^{\mathrm{diag-a}}$&0.012&0.010&0.010&0.020\\ &$J_{ij,i}^{\mathrm{off-s},x}$&0.019&0.007&0.009&0.013\\
&$\vert \vec{D}_{ij}^{x}\vert$& 0.197&0.380&0.618&2.941\\\cline{2-6}
&$J_{ij,i,\mathrm{dip}}^{\mathrm{diag-s}}$&0.023&0.034&0.007&0.011\\
& $J_{ij,i,\mathrm{dip}}^{\mathrm{diag-a}}$&0.070&0.102&0.015&0.033\\
&$J_{ij,i,\mathrm{dip}}^{\mathrm{off-s},x}$& 0.046&0.068&0.013&0.005 \\
&$\vert \vec{D}_{ij,i,\mathrm{dip}}^{x}\vert$&0.0&0.0&0.0&0.0\\\hline
\end{tabular}
\caption{Maximal dipole-dipole contributions for SSC (top) and SLC (bottom) exchange couplings of different materials: bulk Fe, MnGe for Mn atoms at sites $i$ and $j$ ($r_{ij}=0.61$) and three layers of gold on iron (Au on Fe) for Fe atoms at sites $i$ and $j$ ($r_{ij}=0.71$). For the dipole-dipole interactions we use for the anti-symmetric off-diagonal elements the same notation as for DMI-like parameters, i.e. $\vec{D}_{ij(i),\mathrm{dip}}$. From Eq. \eqref{eq:H_dip1} it directly follows that the anti-symmetric off-diagonal elements of both tensors (and hence the DMI and SLC-DMI) vanish for all materials.}
\label{tab:J_iji_dip}
\end{table}

In order to have a complete picture of the spin-lattice interactions in magnetic materials the contribution to the SSC and SLC tensors from the dipole-dipole interaction is considered. Although the dipole-dipole interaction is treated on a classical level, represented by the Hamiltonian
\begin{align}
    H_{\mathrm{dip}}&=-\frac{\mu_0}{4\pi \vert \vec{r}_{ij} \vert^3 } \left[3\,(\vec{m}_i\cdot  \hat{\vec{r}}_{ij})(\vec{m}_j\cdot  \hat{\vec{r}}_{ij}) - \vec{m}_i \cdot \vec{m}_j\right],
    \label{eq:H_dip1}
\end{align}
for two magnetic moments $\vec{m}_i$ and $\vec{m}_j$ at sites $i$ and $j$ separated by a distance vector $\vec{r}_{ij}$, its contribution arises from the (quantum-electro-dynamical) Breit interaction \cite{Breit1932}. Consequently, it can be considered as a consistent addition to the exchange coupling tensors presented in the previous sections. For a ferromagnetic reference system with $\vec{m}_i\parallel \vec{e}_z$ the SSC contribution is given by
\begin{align*}
J_{ij, \mathrm{dip}}^{xx}&=-\frac{\mu_0}{4\pi \vert \vec{r}_{ij} \vert^3 }m_i^{z,0} m_j^{z,0} \big[3\, (\hat{r}_{ij}^x)^2 -1 \big]\\
J_{ij, \mathrm{dip}}^{yy}&=-\frac{\mu_0}{4\pi \vert \vec{r}_{ij} \vert^3 }m_i^{z,0} m_j^{z,0} \big[3\, (\hat{r}_{ij}^y)^2 -1 \big]\\
J_{ij, \mathrm{dip}}^{xy}&=J_{ij, \mathrm{dip}}^{yx} = -\frac{\mu_0}{4\pi \vert \vec{r}_{ij} \vert^3 }m_i^{z,0} m_j^{z,0} \, 3\, \hat{r}_{ij}^x\hat{r}_{ij}^y
\end{align*}
and the SLC contribution, here for a displacement in $\mu=x$ direction, is
\begin{align*}
J_{ij,i,\mathrm{dip}}^{xxx}&=-\frac{3\mu_0}{4\pi }m_i^{z,0} m_j^{z,0} \frac{3r_{ij}^x((r_{ij}^y)^2+(r_{ij}^z)^2)-2(r_{ij}^x)^3}{\vert r_{ij}\vert ^7}\\
J_{ij,i,\mathrm{dip}}^{yyx}&=-\frac{3\mu_0}{4\pi  }m_i^{z,0} m_j^{z,0} \frac{r_{ij}^x\left((r_{ij}^x)^2-4(r_{ij}^y)^2+(r_{ij}^z)^2\right)}{\vert r_{ij}\vert ^7}\\
J_{ij,i,\mathrm{dip}}^{xyx}&=J_{ij,i,\mathrm{dip}}^{yxx}\\
&=\frac{3\mu_0}{4\pi  }m_i^{z,0} m_j^{z,0} \frac{r_{ij}^y\left( 4(r_{ij}^x)^2- (r_{ij}^y)^2-(r_{ij}^z)^2\right)}{\vert r_{ij}\vert ^7}.
\end{align*}
The dipole-dipole contribution is normally considered to be very small and negligible. Exemplary values are shown in Tab. \ref{tab:J_iji_dip}. It can be seen that the dipole-dipole contribution is even larger than the values obtained from the perturbative method for some SSC as well as SLC components for some of the materials considered.\\

To conclude, we have shown that dipole-dipole interactions can make -- depending on the material under consideration -- a significant contribution to the SLC parameters. This is particularly interesting for the simulation of these materials, e.g. via combined spin-lattice molecular dynamics (MD) simulations: When modeling the combined spin  and lattice dynamics, the largest SLC contributions should be taken into account preferentially. Our results show that one has to carefully consider the various contributions for each material. In particular, we have shown that for some materials the dipole-dipole interaction may even play a leading role, as it has already been assumed in spin-lattice MD simulations, e.g. by Aßmann et al. \cite{Assmann2019} or Strungaru et al. \cite{Strungaru2021}.

\section{Numerical Results}

\subsection{SLC Tensors for collinear ferro- and antiferromagnets}
Here, we will analyze the SLC parameters systematically for various materials with different magnetic structures, investigate the role of spin-orbit coupling, and compare the SLC contributions for bulk geometries and free standing monolayers as well as different substrate materials for deposited magnetic films. Furthermore, we will shine light on the role of different elements on the strength of spin-lattice interactions. Finally, we show that the resulting SLC contributions can be connected to modifications of the phonon dispersion due to spin-lattice coupling as proposed by Ref. \cite{Mankovsky2022}. If not stated differently calculations are performed using LDA-DFT with a k-mesh of $2000$ points and $l_{\mathrm{max}}=3$.

\subsubsection{Dependence on Magnetic Structure: FePt vs. CrPt}
\begin{figure}[t]
\centering
\hspace{-0.2cm}
\includegraphics[width=0.46\textwidth]{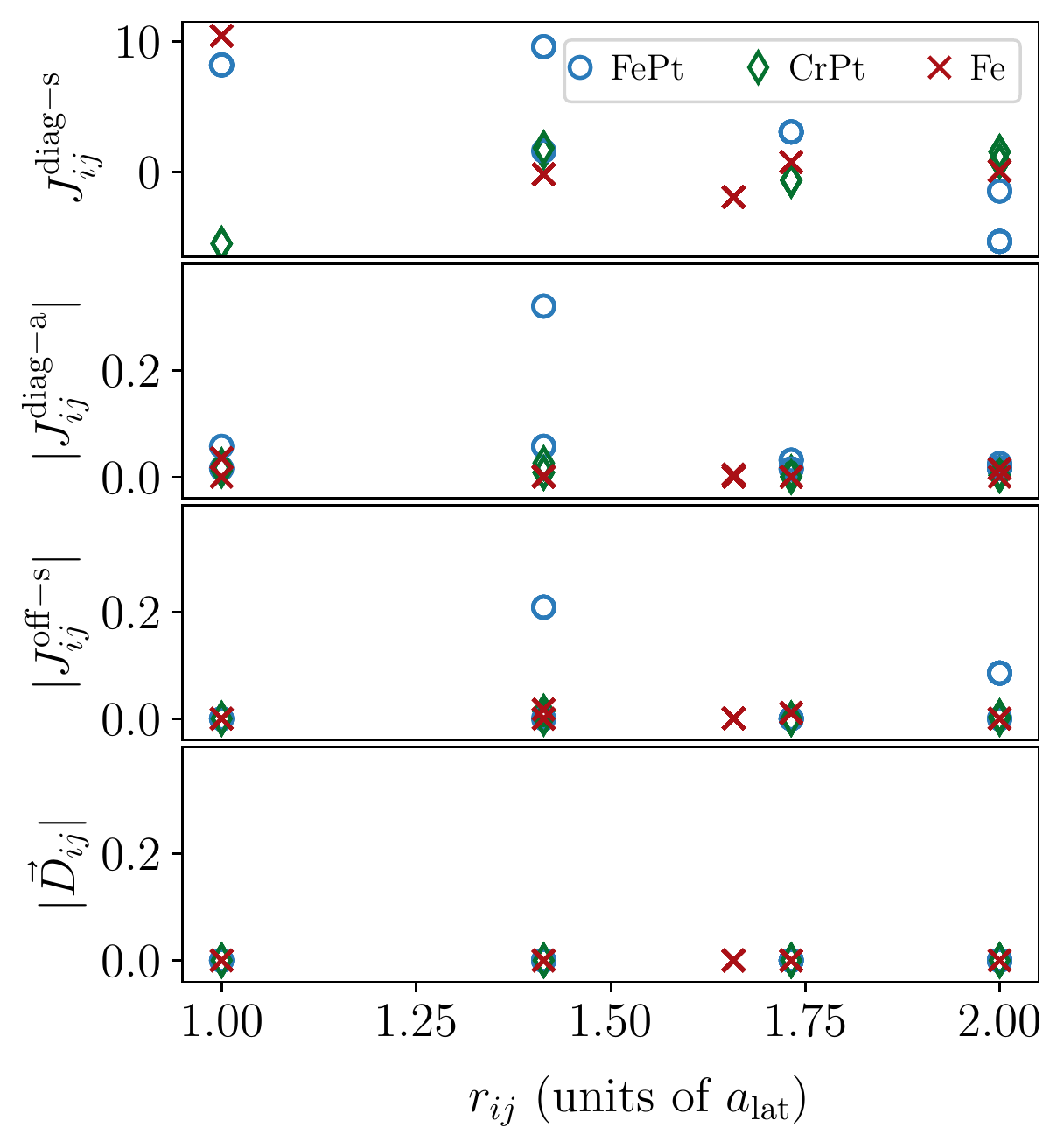}
    \caption{SSC contributions in FePt (circles) and CrPt (diamonds) for atoms $i$ and $j$ being Fe / Cr atoms, compared to the values for bcc Fe (crosses). FePt is ferromagnetic and hence the isotropic part $J_{ij}^{\mathrm{diag-s}}=\frac{1}{2}\left( J_{ij}^{xx}+J_{ij}^{zz}\right)$ is positive. FePt is anisotropic in $z$ directon, i.e. the diagonal anti-symmetric part $J_{ij}^{\mathrm{diag-a}}=\frac{1}{2}\left( J_{ij}^{xx}-J_{ij}^{zz}\right)$ is relatively large. The off-diagonal symmetric part is given by $J_{ij}^{\mathrm{off-s}}=\frac{1}{2}\left( J_{ij}^{xy}+J_{ij}^{yx}\right)$.}
    \label{fig:FePt_CrPt_SSC}
\end{figure}

\begin{figure}[b]
\centering
\begin{minipage}[t]{0.48\textwidth}
    \includegraphics[width=1\textwidth]{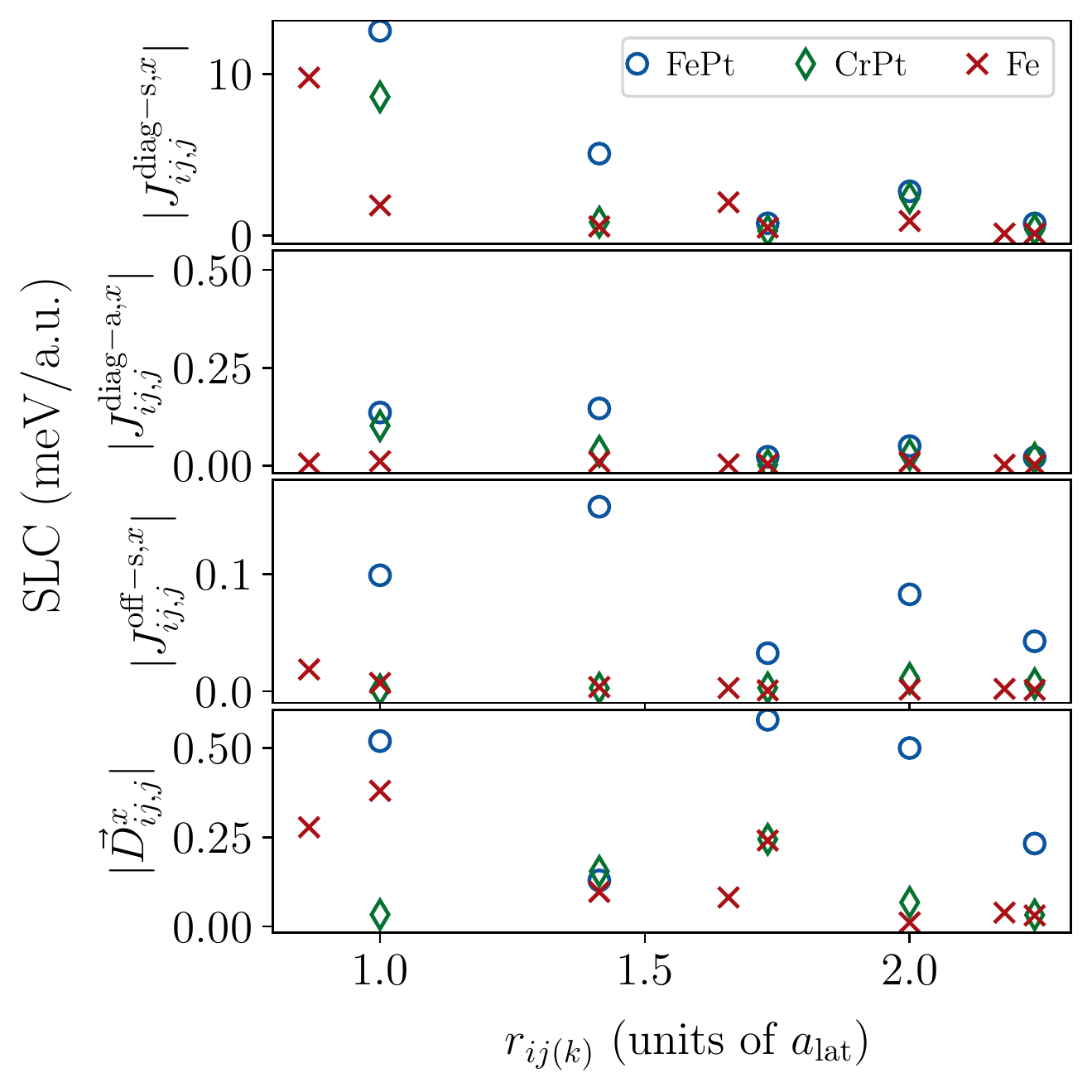}
\end{minipage}
    \caption{Maximal SLC contributions for each distance $r_{ij}$ in FePt (empty circles) and CrPt (filled diamonds) for atoms $i$ and $j$ being Fe / Cr atoms, compared to the values for bcc Fe presented in Ref. \cite{Mankovsky2020} (crosses). FePt as well as CrPt has a large anisotropy in $z$ direction, i.e. the anti-symmetric part of the diagonal components $J_{ij,k}^{\mathrm{diag-a},\mu}=\frac{1}{2}\left(J_{ij,k}^{xx,\mu}-J_{ij,k}^{zz,\mu} \right)$ is relatively large.}
    \label{fig:FePt_vs_CrPt_SLC}
\end{figure}

FePt and CrPt have a very similar lattice structure, both being ordered in the $L1_0$ phase.
 The lattice constants are $a^{\mathrm{FePt}}_{\mathrm{lat}}=2.72$ \AA \mbox{} and $a^{\mathrm{CrPt}}_{\mathrm{lat}}=2.67$ \AA. 
 However, the magnetic properties of both materials are very different. This can be observed in Fig. \ref{fig:FePt_CrPt_SSC}: Whereas FePt is ferromagnetic with strong FM nearest neighbor Fe-Fe interactions, with $J_{ij}^{\mathrm{diag-s}}=8.5\,\mathrm{meV}$ for in-plane neighbors, CrPt is strongly antiferromagnetic with negative nearest Cr neighbor coupling $J_{ij}^{\mathrm{diag-s}}=-4.4\,\mathrm{meV}$ for in-plane neighbors \cite{Schmidt2020}. 
 The induced magnetic moment of Pt is rather small and hence also the respective SSC with Pt at sites $i$ or $j$ are small \cite{Mryasov2005}. 
 Both materials are anisotropic in $z$ direction with a diagonal anti-symmetric part up to $\vert J_{ij}^{\mathrm{diag-a}}\vert=\frac{1}{2}\vert J_{ij}^{xx}-J_{ij}^{zz}\vert=0.32\,\mathrm{meV}$ in FePt and $\vert J_{ij}^{\mathrm{diag-a}}\vert =0.026\,\mathrm{meV}$ in CrPt for out-of-plane neighbors, which is relatively large compared to the DMI and symmetric off-diagonal contributions.
 
 \begin{figure}
\centering
    \begin{minipage}{0.3\textwidth}
    \includegraphics[width=1\textwidth]{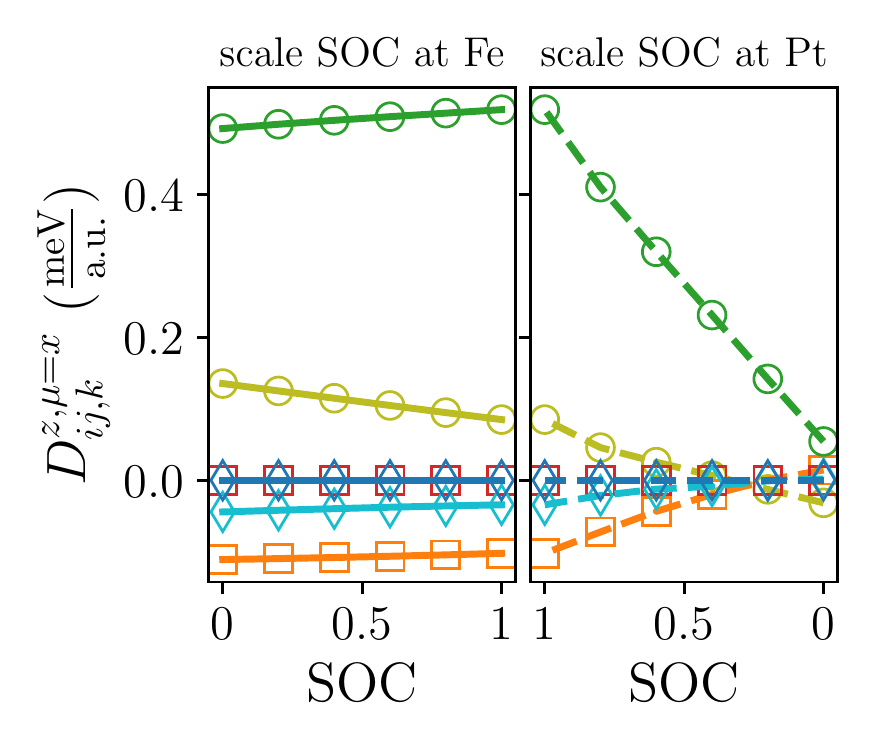}
    \end{minipage}
    \hspace{-1em}\vspace*{-5em}
    \begin{minipage}{0.18\textwidth}
    \includegraphics[width=1\textwidth]{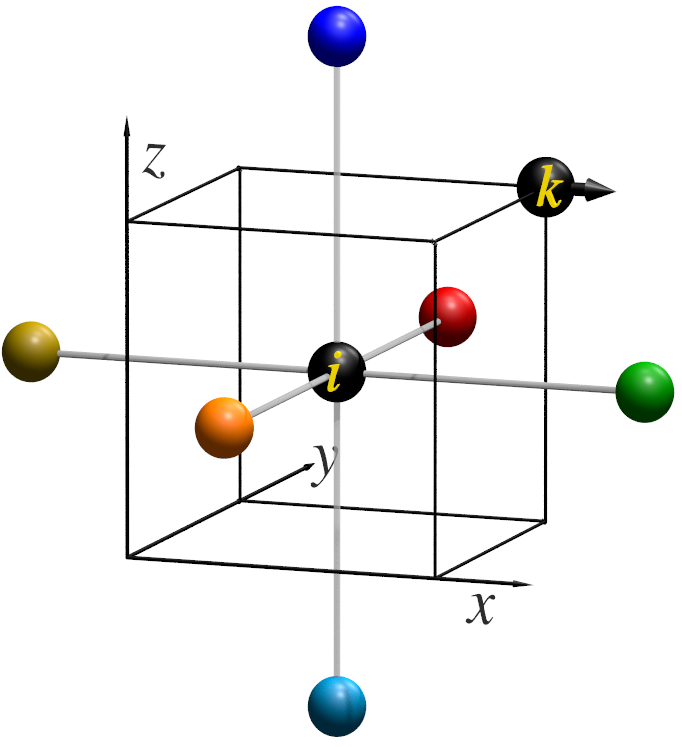}
    \end{minipage}
    \vspace{5em}
    \caption{Dependence of SLC-DMI $D_{ij,k}^{z,\mu=x}$ in FePt for all nearest neighbors on the strength of SOC, scaled at the Fe (left) or Pt (right) sites, for iron atoms at sites $i$ and $j$. The displaced atom is a Pt atom with $r_{ik}=(0.5, 0.5, 0.5)\,a_{\mathrm{lat}}$. }
    \label{fig:FePt_SOC}
\end{figure}

In Fig. \ref{fig:FePt_vs_CrPt_SLC} the SLC parameters for FePt and CrPt are presented. In the case of FePt the isotropic SLC part decays exponentially, starting from $\vert J_{ij,j}^{\mathrm{diag-s},x}\vert=\frac{1}{3}\vert J_{ij,j}^{xx}+J_{ij,jls}^{zz}\vert=12.8\,\mathrm{meV/a.u.}$ for neighboring Fe atoms. The same behavior is observed for CrPt with slightly smaller isotropic parts up to
$\vert J_{ij,j}^{\mathrm{diag-s},x}\vert =8.6\,\mathrm{meV/a.u.}$ for neighboring Cr atoms. Note that $J_{ij,j}^{\mathrm{diag-s},x}$ can be positive and negative for ferromagnetic as well as antiferromagnetic materials, depending on the position of atom $j$ w.r.t. the displaced atom $i$. The second-largest SLC contributions are the on-site anti-symmetric diagonal parts and the DMI contribution (both around $4\,\%$ of the isotropic part) for nearest Fe sites in FePt and the DMI contribution (around $0.6\,\%$ of the isotropic part) in CrPt for nearest Cr neighbors. However, there is a relatively large SLC-DMI contribution for neighbors with $r_{ij}=(0, 1.0, 1.4)\,a_{\mathrm{lat}}$, e.g. with $\vert \vec{D}_{ij,j}^{x}\vert = \frac{1}{2}\vert J_{ij,j}^{yz}-J_{ij,j}^{zy}\vert  = 0.58\,\mathrm{meV/a.u.}$ ($\vert \vec{D}_{ij,j}^{x}\vert = 0.28\,\mathrm{meV/a.u.}$) compared to $\vert J_{ij,j}^{\mathrm{diag-s},x}\vert = 0\,\mathrm{meV/a.u.}$ for the same neighbors or $\vert J_{ij,j}^{\mathrm{diag-s},x}\vert = 0.65\,\mathrm{meV/a.u.}$ ($\vert J_{ij,j}^{\mathrm{diag-s},x}\vert = 0.29\,\mathrm{meV/a.u.}$) for other neighbors with the same distance from atom $i$ for FePt (CrPt). Compared to Fe, the SLC-DMI values are in the same order of magnitude, whereas the anisotropic parts of bcc Fe $J_{ij}^{\mathrm{diag-a}}$ and $J_{ij,j}^{\mathrm{diag-a}}$ are much smaller. Note that here we consider the asymmetry of the diagonal parameters between $x$ and $z$ directions, instead of $x$ and $y$ considered in Ref. \cite{Mankovsky2022}. Hence, it seems that a material with strong magnetic anisotropy (i.e. large $J_{ij}^{\mathrm{diag-a}}$) has an anti-symmetric diagonal part which is more affected by the displacement than for materials with small anisotropy parts.\\

Since the SPRKKR program \cite{SPRKKR} used for the present investigations allows to scale the strength of the relativistic spin-orbit correction, it is possible to investigate the role of the SOC on the atoms which mediate the SLC interaction. This is shown in Fig. \ref{fig:FePt_SOC}. It presents the effect of the SOC on the SLC for nearest Fe neighbors (left) and on pairs of Fe atoms $i$ and $j$ up to $r_{ij}\leq 2\,\,a_{\mathrm{lat}}$ for a lattice distortion at a Pt site $k$ with $r_{ik}=(0.50, 0.50, 1.41)\,a_{\mathrm{lat}}$ (right). It can be seen that the SLC-DMI exhibits only a weak dependence on the strength of the SOC at the Fe sites, but a strong dependence on the SOC of the Pt atoms: The $z$-component of the SLC-DMI decreases by almost one order of magnitude when the SOC on the mediating Pt atom is scaled to zero. This indicates that the mediating atoms play a key role for the influence of the spin-lattice coupling.

\subsubsection{Dependence on Dimensionality: Bulk Iron, Free Standing Iron Monolayer and Substrates}

\begin{table}
\centering
\begin{tabular}{c|c|c|c|c}
SSC &bulk Fe& Fe(001)&Fe(111)&Fe on Ir \\\hline
$J_{ij}^{\mathrm{diag-s}}$ &11.39 &25.29&24.02&7.35 \\
$J_{ij}^{\mathrm{diag-a}}$ &0.00 &0.05&0.03&0.07 \\
$J_{ij}^{\mathrm{off-s}}$ & 0.01&0.00&0.03&0.08 \\
$D_{ij}^z$ &0.00 &0.00&0.00&1.84 \\
\end{tabular}
\caption{Maximal absolute SSC contributions in meV in bulk Fe, free standing monolayers with their surface perpendicular to the [001] and [111] directions and a deposited Fe film on Ir with its surface perpendicular to the [111] direction. Note that the contributions come from different neighbors with different distances, not necessarily nearest neighbors.}
\label{tab:Fe_bulk_monolayers_SSC}
\end{table}

\begin{table}[t]
\centering
\begin{tabular}{c|c|c|c|c|c}
SLC &$\mu$&bulk Fe& Fe(001) &Fe(111) &Fe on Ir \\\hline
\multirow{2}{*}{$J_{ij,j}^{\mathrm{diag-s},\mu}$}&x &9.79 & 18.90&32.09&10.80\\
&y & 9.79& 18.90 & 27.79&10.54\\
&z & 9.79& 0.09&0.29&9.61\\\hline
\multirow{2}{*}{$J_{ij,j}^{\mathrm{diag-a},\mu}$}&x &0.01 & 0.22&0.03&0.15\\
&y &0.01 &0.27 &0.03&0.17\\
&z &0.01 &0.03 &0.0&0.15\\\hline
\multirow{2}{*}{$J_{ij,j}^{\mathrm{off-s},\mu}$}&x & 0.02& 0.05&0.03 &0.14\\
&y &0.02 &0.05&0.04&0.12 \\
&z &0.02 & 0.00&0.0&0.15\\\hline
\multirow{2}{*}{$ D_{ij,j}^{z,\mu}$}&x &0.38 & 0.47&0.28&2.62\\
&y & 0.38& 0.47&0.33&2.28\\
&z &0.04 & 0.00&0.30&0.31 \\\hline
\end{tabular}
\caption{ Maximal absolute SLC contributions for $j=k$ and displacement in $x$ direction in meV/a.u. in bulk Fe, free standing monolayers with their surface perpendicular to the [001] and [111] directions and a deposited Fe film on Ir with its surface perpendicular to the [111] direction.}
\label{tab:Fe_bulk_monolayers_SLC}
\end{table}

Here, we compare the SSC and SLC parameters for materials with different dimensionality. We will compare the exchange interactions for bulk Fe, free standing monolayers with their surface perpendicular to the [001] and [111] directions and substrate materials with a Fe layer deposited on a metal $M(111)$ surface and $M = $ Ir. 
From tables \ref{tab:Fe_bulk_monolayers_SSC} and \ref{tab:Fe_bulk_monolayers_SLC} it can be seen that the dimensionality of the considered material affects both SSC and SLC significantly: For the SSC, the isotropic exchange of the two dimensional free-standing film is more than twice as large as for the bulk material. Apart from that, all other contributions vanish or are much smaller than the isotropic part in both materials. In contrast, the SLC have other significant contributions to the SLC tensor apart from the isotropic exchange. In both materials the SLC-DMI $\vec{D}_{ij,j}^\mu$ is the second largest contribution to the SLC tensor, as reported for bulk Fe by Ref. \cite{Mankovsky2022}. Even when comparing to the SSC contributions, $\vec{D}_{ij,j}^\mu\cdot u_j^\mu$ gives the second largest energy contribution for a realistic displacement of e.g. around $2\%$ of the lattice constant. This term hence can significantly contribute to angular momentum transfer between the spin system and the lattice. An exception is the SLC-DMI contribution for an out-of plane displacement $\vec{D}_{ij,j}^z$ for Fe(001). Furthermore, there is a relatively large SLC anisotropic part $J_{ij,j}^{\mathrm{diag-a},x}$ and $J_{ij,j}^{\mathrm{diag-a},y}$ in the monolayer, but not in the bulk material, which can also be related to spin-lattice angular momentum transfer.  Note that the contributions in tables \ref{tab:Fe_bulk_monolayers_SSC} and \ref{tab:Fe_bulk_monolayers_SLC} come from different neighbors with different distances, not necessarily nearest neighbors.\\

For bulk Fe, displacements in $x$, $y$ and $z$ directions have the same maximal absolute contributions. Note, however, that this symmetry is actually broken since the magnetization direction (here: $z$ direction) is taken into account. This becomes clear when considering the individual tensor elements, with e.g. $J_{ij,j}^{xy,x}=0.22\,\frac{\mathrm{meV}}{\mathrm{a.u.}} \neq J_{ij,j}^{xy,z}=0.02\,\frac{\mathrm{meV}}{\mathrm{a.u.}}$. In the monolayer, where the magnetization is aligned in the out-of-plane direction, the asymmetry of $x$ / $y$ directions compared to the $z$ direction is much larger and can also be observed in the maximal SLC tensor contributions in Tab. \ref{tab:Fe_bulk_monolayers_SLC}.\\

\begin{figure}[t]
\centering
    \hspace{-0.5cm}
   \includegraphics[width=0.23\textwidth]{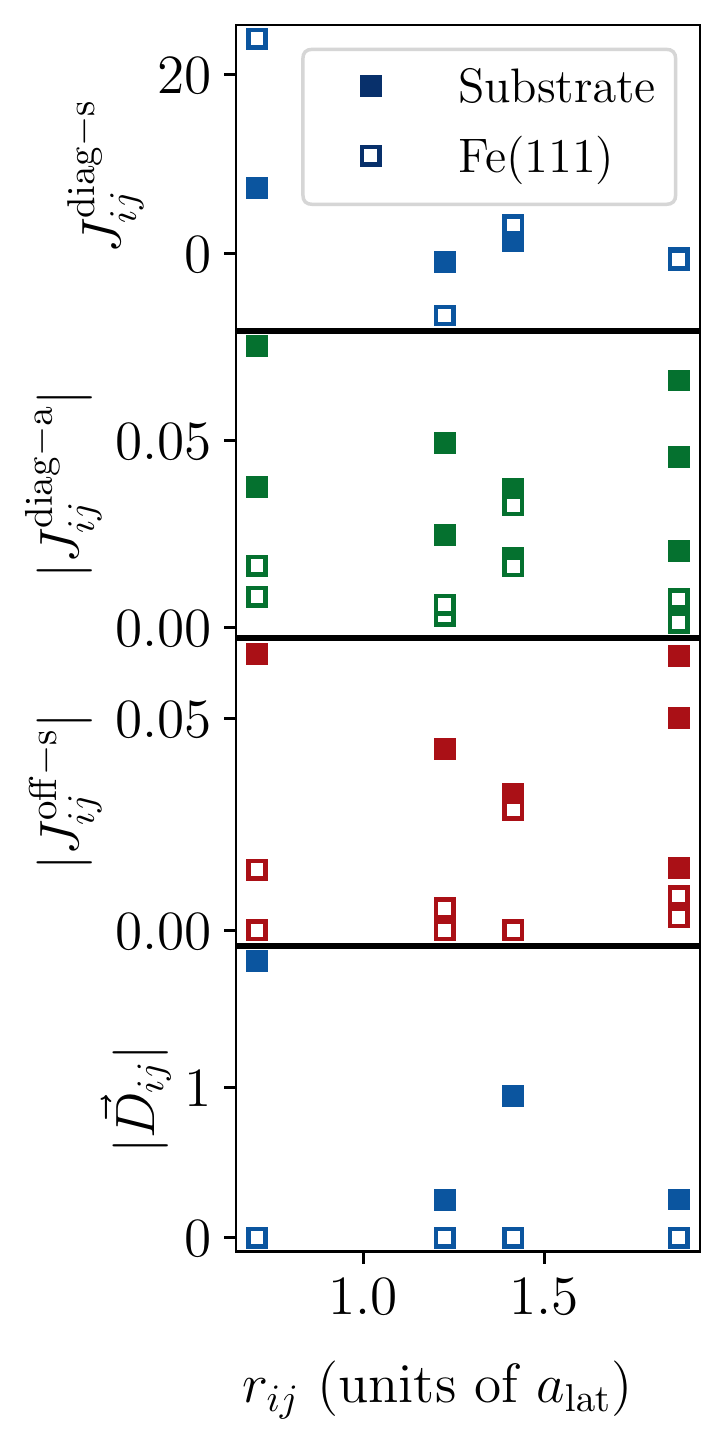}
   \hspace{-0.3cm}
    \vspace{-0.1cm}
   \includegraphics[width=0.252\textwidth]{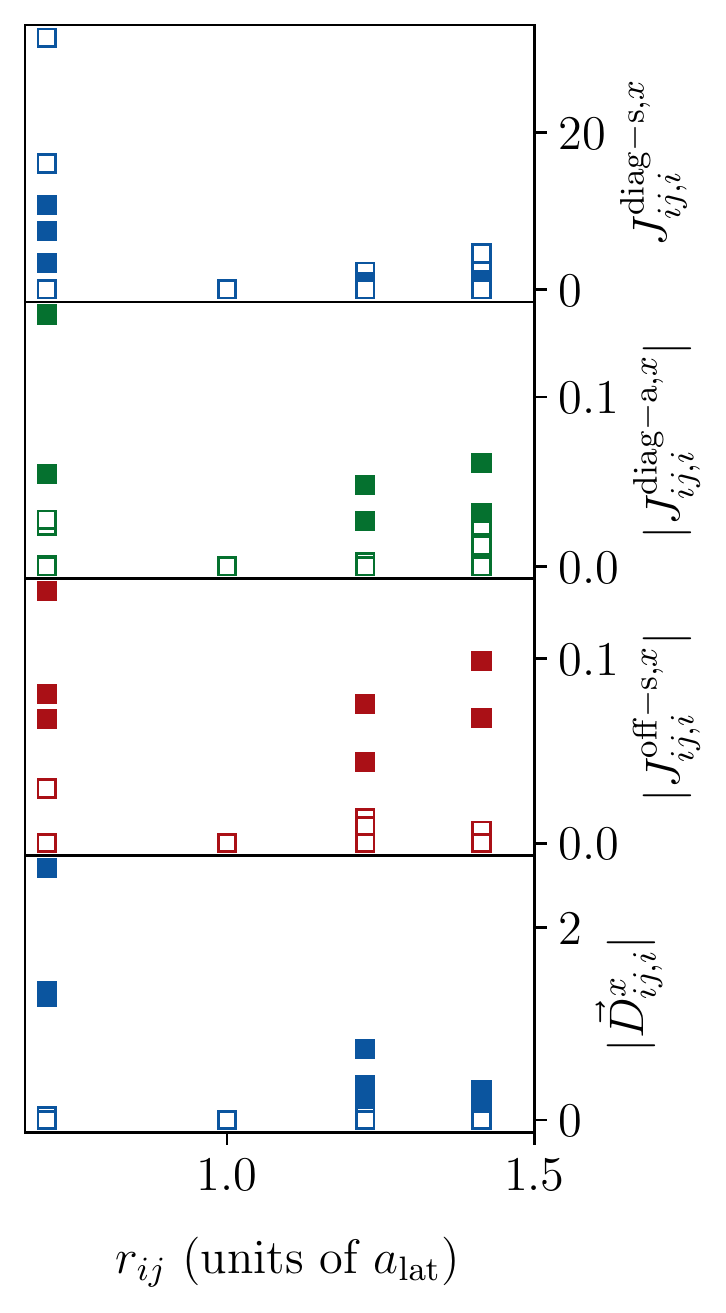} 
    \caption{Comparison of SLC contributions for a surface layer system consisting of a Fe monolayer on Ir$(111)$ and a free standing Fe(111) monolayer. Sites $i$ and $j$ are occupied by a Fe atom.}
    \label{fig:Substrates}
\end{figure}

The effect of the dimensionality of the considered system on the SSC and SLC contributions can also be observed when investigating the transition between 2D monolayers to substrate system consisting of a Fe monolayer with its surface perpendicular to (111) on three layers of a metal (here Ir). The respective spin-spin and spin-lattice exchange coupling contributions for $i$ being a Fe atom and $i=k$ are shown in tables \ref{tab:Fe_bulk_monolayers_SSC} and \ref{tab:Fe_bulk_monolayers_SLC} as well as in Fig. \ref{fig:Substrates}. Further details on the calculations can be found in the next section. It can be seen that the results for the free standing monolayer and the substrate materials strongly differ in the spin-spin and spin-lattice case. For the SSC and SLC isotropic parts, the monolayer has stronger exchange contributions for most configurations of $i$ and $j$ for the SSC and SLC. In contrast, other contributions related to anisotropy and angular momentum transfer \cite{Mankovsky2022}, are dominated by the substrate material for SSC and SLC and all distances between atoms $i$ and $j$. In particular, there is a finite $z$ component of the SSC-DMI which is related to the lack of inversion symmetry in the substrate. For the monolayer, $D_{ij}^z=0$ as already presented in Tab. \ref{tab:Fe_bulk_monolayers_SSC}. Also in the SLC-DMI case the contributions of the substrate dominate.

\subsubsection{Dependence on Element Types: Substrates}
So far we focused on the properties of the SLC parameters
$J_{ij,k}$ assuming $k=j$, characterizing the modification of
the exchange interactions between the atoms on sites $i$ and $j$, when
one of the atoms is displaced. This however does not imply that an
impact of displacements of the atoms on sites $i\neq k \neq j$ can be
neglected, although the role of these displacements depends on the
material, in particular on the origin of the exchange interaction. To discuss this contribution we consider
2D materials consisting of a Fe monolayer (ML) deposited on a $M(111)$ surface and investigate the role of the type of substrate $M=\mathrm{Ir},\, \mathrm{Pt},\, \mathrm{Au}$. The elements Ir, Pt and Au belong to the sixth period of the periodic table and to neighboring groups $9$, $10$ and $11$. Their electronic configurations are $\left[\mathrm{Xe}\right]4\mathrm{f}^{14} 5\mathrm{d}^{7} 6\mathrm{s}^2$ (Ir), $\left[\mathrm{Xe}\right]4\mathrm{f}^{14} 5\mathrm{d}^{9} 6\mathrm{s}^1$ (Pt) and $\left[\mathrm{Xe}\right]4\mathrm{f}^{14} 5\mathrm{d}^{10} 6\mathrm{s}^1$ (Au). We model the system by a 3ML slab, i.e. calculations were performed for a supercell consisting of two vacuum layers, one Fe surface layer and three $M$ layers representing the substrate. Hence, the considered system actually consists of an infinite stack of substrate layers, separated by two vacuum layers. The number of separating vacuum layers between the surface layers is sufficiently large and the interlayer interactions sufficiently small to consider the system as an isolated surface layer-substrate system. 

\begin{table}
\centering
\begin{tabular}{c|c|cccc|}
displ. $\mu$&$M$& $D_{ij,i}^{x,\mu}$& $D_{ij,i}^{y,\mu}$& $D_{ij,i}^{z,\mu}$&$\vert \vec{D}_{ij,i}^{\mu}\vert$\\\hline
\multirow{3}{*}{x}&Ir &0.49 &0.36 &2.62&2.69 \\
&Pt & 1.23&0.05 &0.55 &1.35  \\
&Au &0.99 &0.44 &2.76 &2.97\\\hline
\multirow{3}{*}{y}&Ir  &0.31 &0.68 & 2.25& 2.37\\
&Pt & 0.60 &1.15 &0.69 &1.46 \\
&Au &1.21 &0.22 &3.07 &3.30\\ \hline
\multirow{3}{*}{z}&Ir &1.77& 4.03&0.31 &4.41\\
&Pt  &0.11 &0.98 &0.38 &1.06 \\
&Au &0.62 &3.56 &0.04 &3.61 
\end{tabular}
\caption[SLC-DMI in three layers of $M=\mathrm{Ir},\, \mathrm{Pt},\, \mathrm{Au}$ on Fe]{Absolute values of SLC-DMI in three layers of $M=\mathrm{Ir},\, \mathrm{Pt},\, \mathrm{Au}$ on Fe for displacements $\mu$ and a Fe atom at sites $i$ and $j$.}
\label{tab:J_iji_Substrates}
\end{table}
The maximal nearest-neighbor SLC-DMI values for other substrate materials with substrates $M=$ Ir, Pt and Au are shown in Tab. \ref{tab:J_iji_Substrates}. It can be seen that the substrates $M=\mathrm{Ir}$ and $M= \mathrm{Au}$ have a SLC-DMI with similar absolute values and the same largest components for all displacements. For Pt, the results are in general much smaller, with different largest SLC-DMI components.

\begin{figure}[t]
\centering  \includegraphics[width=0.34\textwidth]{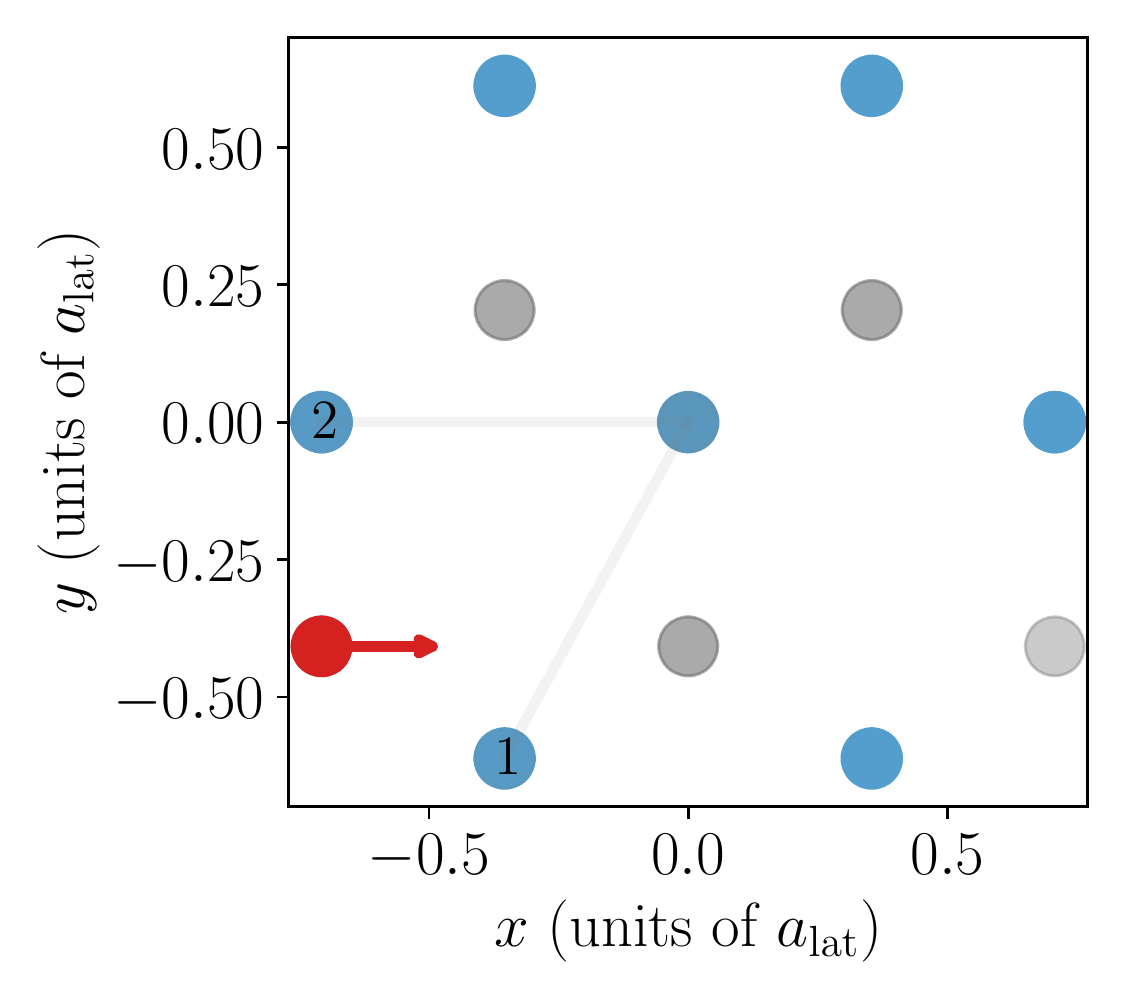}
    \caption{SLC contributions for a surface layer system consisting of a Fe monolayer on Ir$(111)$. Fe atoms are represented by blue dots with the central atom being atom $i$ and atoms 1 (2) being $j$.The red dot and grey dots represent the displaced and the other Ir atoms in the first substrate layer. }   \label{fig:Substrates_SLC_ijk}
\end{figure}

The contribution of SSC and SLC with Ir atoms as one of the interacting atoms (e.g. atom $j$ in Fig. \ref{fig:Substrates}) are zero due to the vanishing magnetic moment of Ir atoms. However, the configuration of the Ir atoms can strongly affect the spin-lattice exchange strength. This is visualized in Fig. \ref{fig:Substrates_SLC_ijk} for a specific configuration with an Ir atom (red) located in the first Ir layer on the Fe layer and displaced in $x$ direction. Here, the diagonal SLC contributions are not negligible, i.e. $J_{ij,k}^{\mathrm{diag-s}}(\vec{r}_1)=3.30 \,\frac{\mathrm{meV}}{\mathrm{a.u.}}$ and $J_{ij,k}^{\mathrm{diag-s}}(\vec{r}_2)=-0.40 \,\frac{\mathrm{meV}}{\mathrm{a.u.}}$. For the $z$ component of the SLC-DMI we find $D^z_{ij,k}(\vec{r}_1)=0.16\, \frac{\mathrm{meV}}{\mathrm{a.u.}}$ and $D^z_{ij,k}(\vec{r}_2)=0.39\, \frac{\mathrm{meV}}{\mathrm{a.u.}}$.\\

\FloatBarrier
\subsection{Spin-Lattice Effects in Frustrated Antiferromagnets \label{sec:AFM}}
As seen in the previous section, the magnetic properties determined by the spin-spin exchange interaction tensor $J_{ij}$ can change significantly when displacing atoms $i$ and $j$, or a third atom $k\neq i,j$. This becomes particularly interesting in frustrated antiferromagnets where the magnetic configuration depends extremely sensitively on changes of $J_{ij}$ for neighbors $j$ in different directions.

To discuss this issue, we consider in this section the spin-lattice
interactions for CuCrO$_2$ \cite{POL+16,POL+16} which belongs to
a family of triangular lattice antiferromagnets (TLA) $A$CrO$_2$, exhibiting
interesting magnetic and magnetoelectric and magnetoelastic properties
determined by geometrical spin frustration \cite{Engelsman1973,
  Rasch2009, Carlsson2011, Kimura2003, Singh2009}.
This compound is characterized by a leading AFM nearest-neighbor (nn)
Cr-Cr interaction and weak interactions for an increasing distance
between the interacting Cr atoms.
As is discussed in the literature, the nn Cr-Cr
interactions in these materials may be treated in terms of two
competing contributions \cite{Angelov1991, Delmas1978, Hewston1987, Mazin2007, 
  Rasch2009, Ushakov2013}: 
(a) direct antiferromagnetic interactions of neighboring Cr atoms and 
(b) indirect ferromagnetic superexchange interactions mediated by O
atoms (since $\mathrm{Cr}_1-\mathrm{O}- \mathrm{Cr}_2$ form an angle
of $\approx 90^\circ$), indicating that the leading contribution is the
direct Cr-Cr interaction. 
As a consequence, different SLC parameters can be contributed by
different types of exchange interactions (see Appendix \ref{sec:spit-SLC}).

Considering the diagonal symmetric SLC parameters $J^{\mathrm{diag-s},\mu}_{ij,k} =
\frac{1}{2}(J^{xx,\mu}_{ij,k} + J^{yy,\mu}_{ij,k})$, the parameters with
$k$ characterizing the position of an atom O can be associated mainly with the
superexchange mechanism since a displacement of atom O does not
change the Cr-Cr distance. On the other hand, the
$J^{\mathrm{diag-s},\mu}_{ij,j}$ parameters are connected first of all
with the direct exchange. 

\begin{figure}
\centering
\includegraphics[width=0.49\textwidth,angle=0,clip]{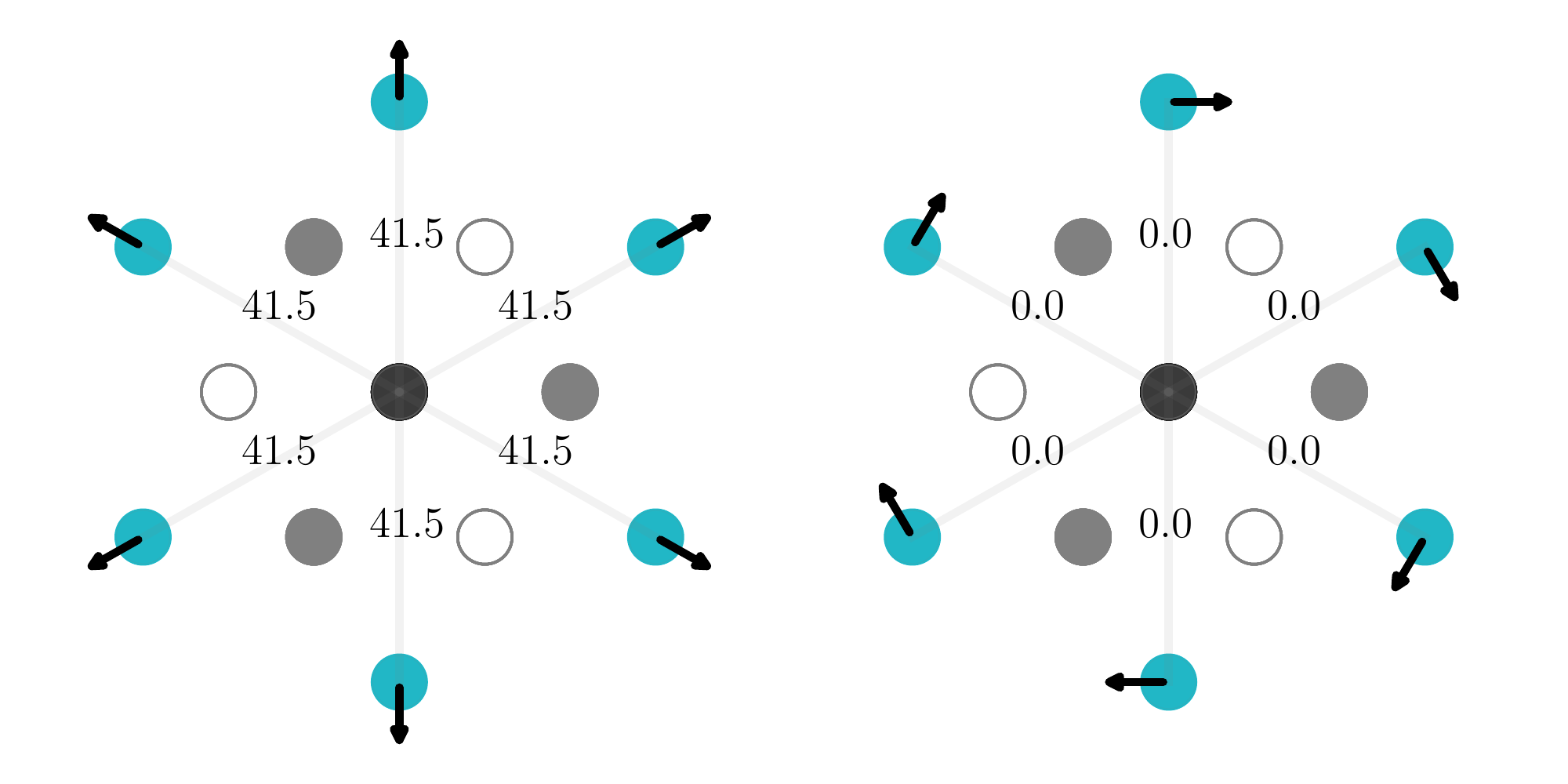}\,
\caption{Longitudinal $J_{ij,j}^{\mathrm{diag-s},lng}$ (left) and
  transverse $J_{ij,j}^{\mathrm{diag-s},trans}$ (right) SLC parameters
  for$\mathrm{CuCrO}_2$, 
  defined in Eqs.\ \eqref{eq:Proj_long} and \eqref{eq:Proj_trans} in
  meV/a.u. }  
  \label{fig:CuCrS2-SLC-Cr_longi_trans}    
\end{figure}

First we will discuss the properties of the SLC parameters
$J^{\mathrm{diag-s},\mu}_{ij,j} =
\frac{1}{2}(J^{xx,\mu}_{ij,j} + J^{yy,\mu}_{ij,j})$ corresponding to $k
= j$. They characterize the change of the Cr-Cr exchange interaction due to 
displacement of one of the interacting Cr atoms, or 
alternatively, they may be seen as the parameters characterizing 
the force $\vec{f}_{ij,j}$ acting on atom $j$, induced by spin
tiltings on sites $i$ and $j$, having the components  $f^\mu_{ij,j}
\sim J^{\mathrm{diag-s},\mu}_{ij,j}$. 
It is instructive to represent the SLC parameters in terms of
\enquote{longitudinal} and \enquote{transverse} displacements,
or \enquote{longitudinal} and \enquote{transverse} induced forces.
The \enquote{longitudinal} force on atom $j$ relative to atom $i$
is oriented along the $\hat{e}_{ij} =  \vec{r}_{ij} / \vert
\vec{r}_{ij}\vert$ direction, that gives
\begin{eqnarray}
{J}^{\mathrm{diag-s},lng}_{ij,j} &=& {J}^{\mathrm{diag-s},x}_{ij,j}(\hat{x} \cdot
\hat{e}_{ij}) + {J}^{\mathrm{diag-s},y}_{ij,j} (\hat{y} \cdot
\hat{e}_{ij}).
\label{eq:Proj_long} 
\end{eqnarray}
The \enquote{transverse} forces perpendicular to this direction are
characterized by corresponding SLC parameters given by
\begin{eqnarray}
 {J}^{\mathrm{diag-s},trans}_{ij,j} &=& {J}^{\mathrm{diag-s},x}_{ij,j} (\hat{x} \cdot
[\hat{e}_{ij} \times \hat{z}]) \notag \\
& & \quad + {J}^{\mathrm{diag-s},y}_{ij,j}(\hat{y} \cdot
[\hat{e}_{ij} \times \hat{z}] ) \nonumber \\
  &=& {J}^{\mathrm{diag-s},x}_{ij,j} (\hat{y} \cdot
\hat{e}_{ij}) - {J}^{\mathrm{diag-s},y}_{ij,j}(\hat{x} \cdot
\hat{e}_{ij}).
     \label{eq:Proj_trans} 
\end{eqnarray}
These nearest-neighbor \enquote{longitudinal} (left) and
\enquote{transverse} (right) SLC parameters for 
$\mathrm{CuCrO}_2$ are shown in
Fig. \ref{fig:CuCrS2-SLC-Cr_longi_trans}.
As one can see, the \enquote{longitudinal} SLC are finite and the same for all nearest
neighbors, while the \enquote{transverse} SLC parameters are equal to
zero. This is a consequence of the symmetry of the system
including a 3-fold rotation axis as well as $\sigma_{xz}$ and
$\sigma_{yz}$ mirror planes.
This implies that the forces on atoms $j$ in
$\mathrm{CuCrO}_2$, induced by spin tiltings on
nearest-neighbor Cr sites via ${\cal J}^{\mathrm{diag-s},\mu}_{ij,j}$
interactions are oriented along the lines connecting these two atoms.
%
\begin{figure}[t]
\includegraphics[width=0.46\textwidth,angle=0,clip]{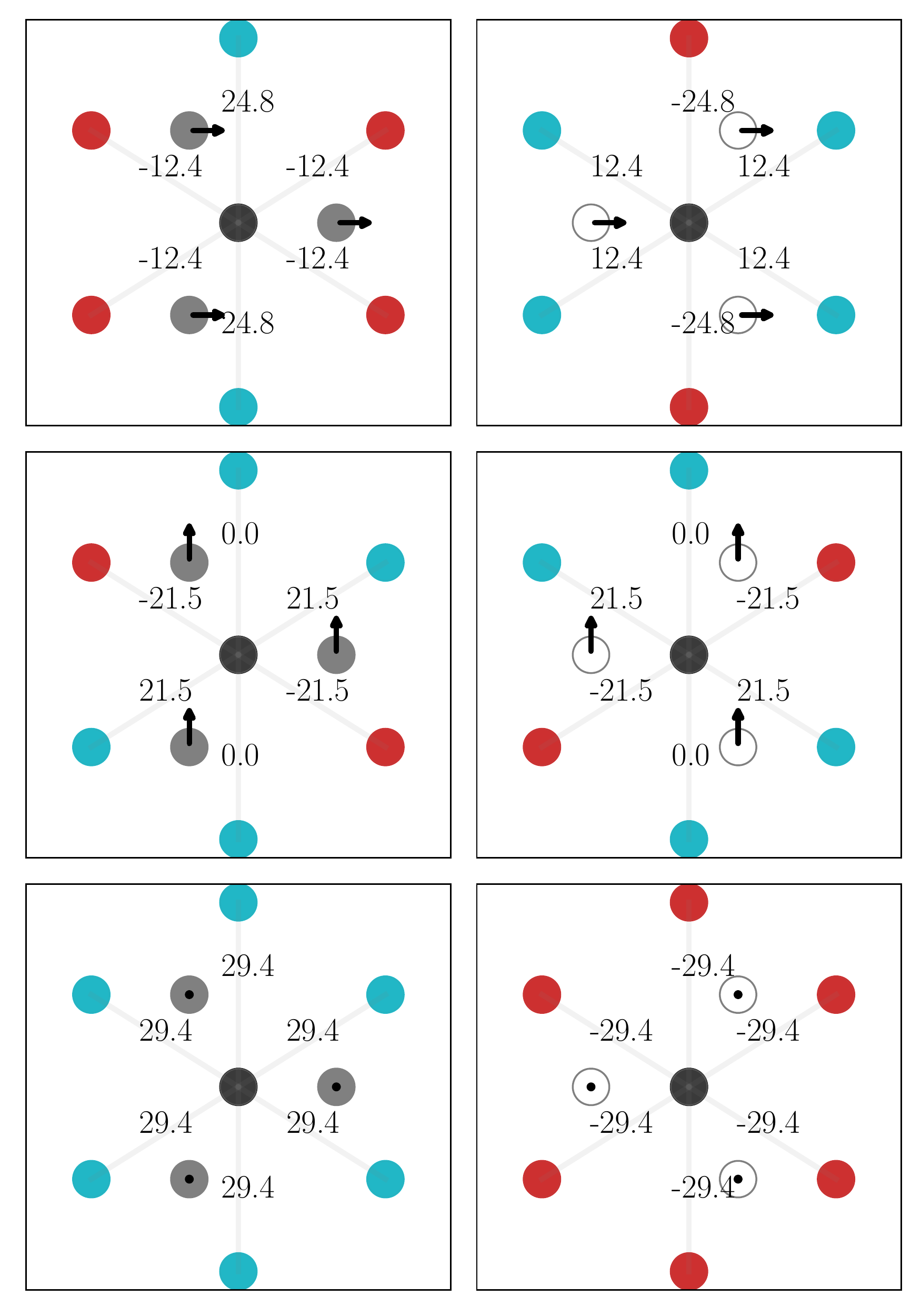}
\caption{ Symmetric SLC $J_{ij,k}^{\mathrm{diag-s},\mu}$ in $\mathrm{CuCrO}_2$ for Cr sites at $i$ and $j$ and for a displacement in $\mu=x,\,y,\,z$ directions (first, second, third row) at a O site $k$ in the layer below (solid dots, left column) and above (empty dots, right column) the Cr layer (in meV/a.u.).
}
\label{fig:CuCrO2-SLC-O}
\end{figure}

As a next step, we discuss the three-site SLC parameters ${\cal 
  J}^{\mathrm{diag-s},\mu}_{ij,k}$ corresponding to nearest-neighboring Cr
atoms in the positions $i$ and $j$ and the O atom on site $k$, which originate
from the superexchange mechanism (Appendix \ref{sec:spit-SLC}). 
These parameters are presented in figure \ref{fig:CuCrO2-SLC-O} for $\mu
= \{x,y,z\}$. Filled gray circles in the left column correspond to the
O layer below and empty circles in the right column correspond to the
O layer above the Cr layer in $\mathrm{CuCrO}_2$. These parameters have
the same order of magnitude as the SLC related to the displacement of
one of the Cr atoms at $k=j$. Again, one can observe the impact of the
crystal symmetry on the SLC parameters, which yields different values
for atoms $k$ located above and below the Cr layer.
This contribution may be important since it represents
  the impact of the spin-lattice coupling on the phonon spectra,
  in addition to the local spin-lattice term \cite{ENR+76,AG71} describing the 
  interplay of spin degree of freedom with the displacements of
  the non-magnetic atoms.

In summary, we can see comparable values of the SLC parameters ${\cal 
  J}^{\mathrm{diag-s},\mu}_{ij,k}$ in the cases of  $k = j$ and  $k \neq
j$, indicating in general the same significance of both of them
for a possible lattice distortion or phonon modes modification
concomitant to magnetic ordering in the system.

To complete the discussion, we represent also the 
properties of the  DMI-like SLC parameters for CuCrO$_2$.
The ${\cal D}^{\alpha,\mu}_{ij,k}$  components are presented in
Tab. \ref{tab:table_CuCrO2_overview} for three different directions of
displacement $x, y, z$, respectively for the case $k=j$ (i.e. $k$ site
occupied by Cr (top)) and  $k \neq j$ (i.e. $k$ site 
occupied by O (bottom)). For convenience, for every pair of nearest neighbor Cr atoms
$i$ and $j$, one can consider displacements of the O  atom within the
corresponding planes, either along the $\vec{R}_{ij}$ vector or
perpendicular to it. Therefore, in both cases, top and bottom, it is
sufficient to see the properties of the DMI-like parameters between the
Cr atoms connected by vector $\vec{R}_{ij} = (0,a,0)$
(i.e. $\hat{r}_{ij} = \hat{y}$ ). 
\begin{table}[t]
  \begin{center}
    \begin{tabular}{|p{2cm}|p{1.3cm}|p{1.3cm}|p{1.3cm}|}
      \hline
  {$\mu$}   &  {$x$ } &  {$y$ }  & {$z$ } \\    \hline
            & Cr         &  Cr         & Cr       \\
      \hline
 ${D}^{x,\mu}_{ij,j}$   & -0.35  & 0 & -0.15  \\
 ${D}^{y,\mu}_{ij,j}$   & 0.0   & -0.04 & 0.0 \\
 ${D}^{z,\mu}_{ij,j}$   & -0.08  & 0.0 &  -0.79  \\
      \hline
    \end{tabular}    
    \begin{tabular}{|l|c|c|c|c|c|c|}
      \hline
  {$\mu$}   & \multicolumn{2}{c|} {$x$ (a)} & \multicolumn{2}{c|} {$y$ (b)}  &
\multicolumn{2}{c|}{$z$ (c)} \\  
      \hline
    & O$_{top}$(l) &  O$_{bot}$ (r) &  O$_{top}$ (l) & O$_{bot}$
(r)&
        O$_{top}$ (l)
                      &  O$_{bot}$ (r)     \\
      \hline
${D}^{x,\mu}_{ij,k} $   & -0.26  &-0.26 & 0 & 0 & -0.21 & -0.21  \\
${D}^{y,\mu}_{ij,k} $   & 0   & 0 & 0.012 & 0.012 & 0 & 0 \\
${D}^{z,\mu}_{ij,k} $   & -0.06  & -0.06 & 0 & 0 & 0.64  &  0.64  \\
      \hline
    \end{tabular}
\vspace{0.5cm}\\

\caption{\label{tab:table_CuCrO2_overview} The DMI-like SLC parameters for CuCrO$_2$
  ${D}^{\alpha,\mu}_{ij,k}$: with the site $k=j$, i.e. occupied by Cr
  (top) and  $k \neq j$, i.e. occupied by O (bottom), for three different directions of 
displacement of the nearest neighbor Cr and O atoms, i.e. $x, y, z$ (see Fig. \ref{fig:CuCrO2-SLC-O}). } 
  \end{center}
\end{table}
Again, these parameters can be seen as a measure for the forces on atoms $j$ (top)
and $k$ (bottom), induced by spin tiltings via different components of
DMI-like SLC.  
As in the case of diagonal symmetric SLC parameters, one can see that
${D}^{\alpha,\mu}_{ij,k}$ are comparable in the cases of $k=j$ and $k\neq j,i$ (where the $k$
site occupied by the O atom), while the DMI-like parameters are smaller
by about two orders of magnitude.

\section{Summary}
To conclude, we have provided a systematic analysis of spin-lattice interactions in collinear ferromagnets and antiferromagnets as well as frustrated AFMs. It was demonstrated that the crystal structure, the magnetic configuration and the dimensionality of the material under consideration determine the strength of the coupling between the spins and the lattice. Furthermore, relativistic effects give rise to non-vanishing components of the spin-lattice coupling tensor that can be connected to a two-site spin-lattice anisotropy and a spin-lattice Dzyaloshinskii–Moriya interaction. Our results are particularly interesting for modeling magnetic materials via combined spin-lattice molecular dynamics simulations since we show that different contributions to the SLC tensor can be relevant depending on the material under consideration. For some materials, even dipole-dipole interactions may have a significant impact on the SLC parameters.

By calculating the respective spin-lattice interactions for CuCrO$_2$ compounds, it was demonstrated that the modification of the spin-spin counterparts is significantly large during the structural transition in these systems and hence might give insights into the magnetic transition that happens simultaneously. The results are benchmarked against calculations performed with embedded clusters, which is so far the most accurate scheme to calculate spin-lattice interactions from first principles.\\



\emph{Acknowledgements.--}
The work in Konstanz was supported by the DFG via project 290/5-2.

\bibliography{main.bib}

\bibliographystyle{apsrev4-1}

\newpage~

\appendix
\section*{Appendix}

\section{SLC for Next-Nearest Neighbors \label{appendix:NNN_EC}}
In this section the results for the change of the diagonal and off-diagonal components of the exchange coupling tensor are presented for next-nearest neighbors $j$ of atom $i$ with $\vert \vec{r}_{ij}\vert =1.0\, a_{\mathrm{lat}}$. In contrast to the nearest neighbors, the investigation of the next-nearest neighbors has the feature that they lie in the $x$, $y$ or $z$ directions and not in a diagonal direction. Hence, the dependence of the modification of $J_{ij}^{\alpha\beta}$ on the direction of the distance vector $\vec{r}_{ij}$ w.r.t. the displacement can be accessed directly.\\

\begin{figure}[ht]
    \centering
    \includegraphics[width=0.46\textwidth]{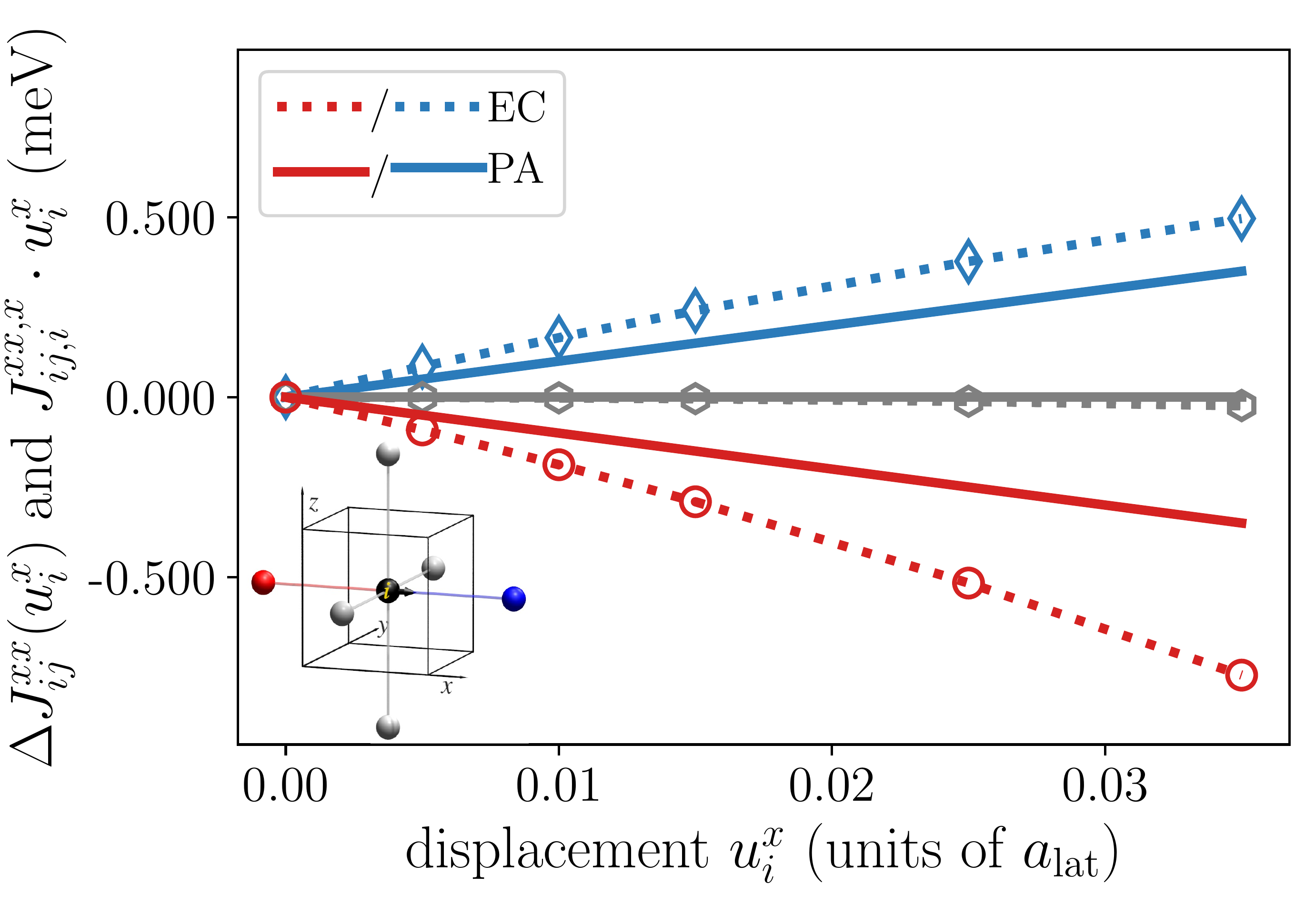}
    \caption{Modification of the diagonal exchange couplings of next-nearest neighbors for embedded clusters (EC) with 27 atoms (dotted lines) and the perturbative method (PA, solid lines) for a displacement in $x$ direction in bcc Fe. The color-code of the atoms $j$ is visualized in the inset figure: Red circles correspond to neighbors with $r_{ij}^x < 0$, blue diamonds to $r_{ij}^x > 0$ and grey hexagons to $r_{ij}^x = 0$.}
    \label{fig:Cluster_Plot12}
\end{figure}

In Tab. \ref{tab:SLC_4} the results for the three-site SLC parameters calculated from $27$ an $51$ atomic clusters for a displacement of $0.005\,a_{\mathrm{lat}}$ in the $x$ direction are compared to the perturbative approach. As for the nearest neighbors, the parameters for different cluster sizes agree within an accuracy up to the second digit. However, in contrast to the nearest neighbor case we find larger differences in the magnitudes of the embedded cluster and perturbative parameters. Nevertheless, the results agree quantitatively and show leading SLC contributions between the same next-nearest neighbors.

\begin{figure}
    \centering
    \includegraphics[width=0.46\textwidth]{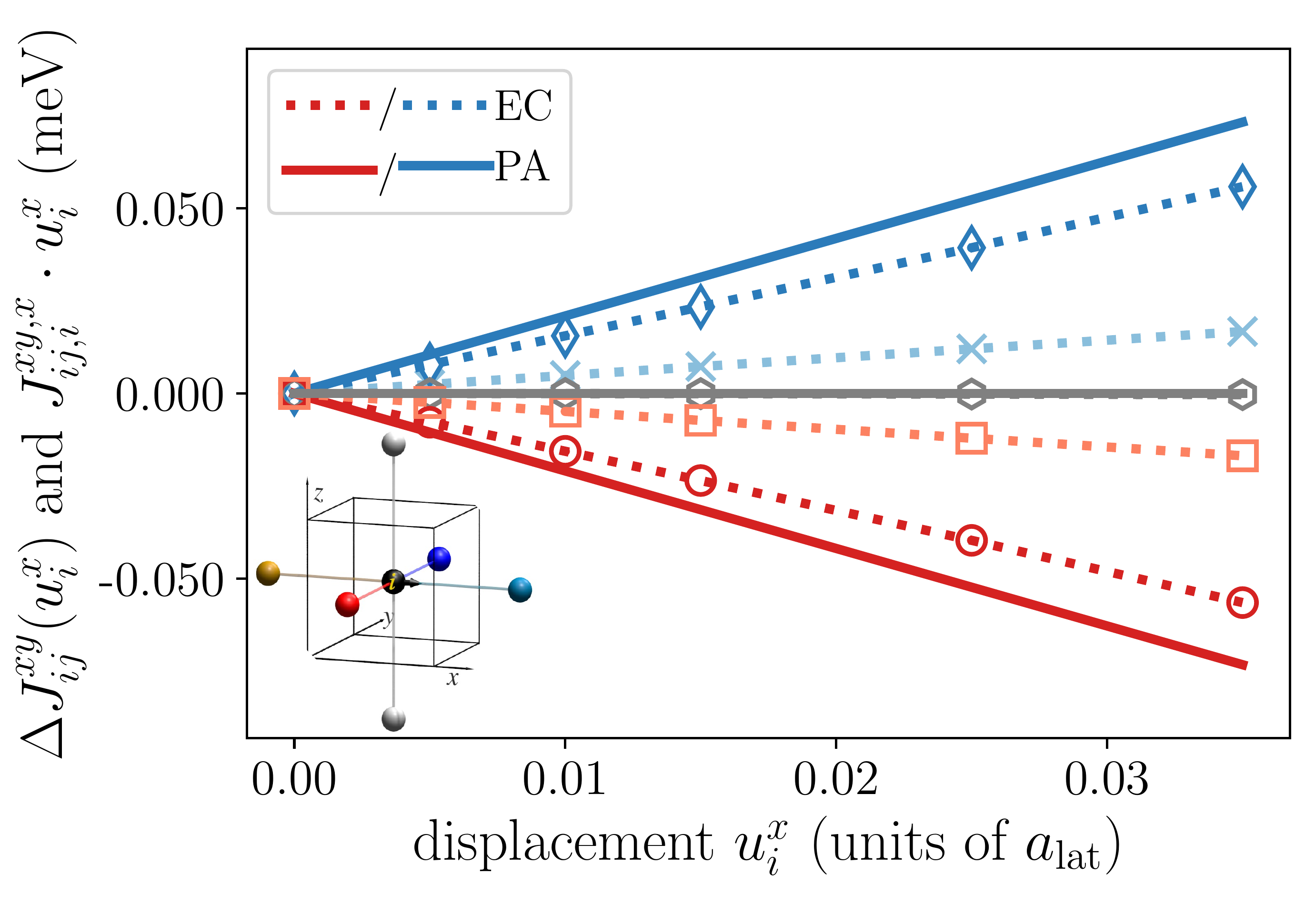}
    \caption{Modification of the off-diagonal exchange couplings of next-nearest neighbors for embedded clusters (EC) with 27 atoms (dotted lines) and the perturbative method (PA, solid lines) for a displacement of one atom in $x$ direction for nearest neighbors in bcc Fe. The color-code of the atoms $j$ is visualized in the inset figure: Dark red circles correspond to neighbors $j$ with $\vec{r}_{ij}=(0, -1, 0)$, light red squares to $\vec{r}_{ij}=(-1, 0, 0)$, dark blue diamonds to $\vec{r}_{ij}=(0,1, 0)$, light blue crosses to $\vec{r}_{ij}=(1, 0, 0)$ and grey hexagons to $r_{ij}^x = 0$. }
    \label{fig:Cluster_Plot22}
\end{figure}

The isotropic part shown in Fig. \ref{fig:Cluster_Plot12} is mainly determined by the next-nearest neighbors $j$ with $\vec{r}_{ij}\parallel \vec{u}_i$. In contrast, the off-diagonal contributions have the largest contributions from $\vec{r}_{ij}\perp \vec{u}_i$, e.g. for a displacement $u_i^x$ in the perpendicular direction $\vec{r}_{ij}^y$. This can be seen in Fig. \ref{fig:Cluster_Plot22}. Furthermore, one can already observe that $J_{ij,i}^{xy,x}\neq J_{ij,i}^{yx,x}$ (the latter is not shown here), which gives rise to an anti-symmetric (Dzyaloshinskii–Moriya) interaction already discussed in Ref. \cite{Mankovsky2022}.

\begin{table}
\centering
\begin{tabular}{c|cc|c}
 &$27$ atoms EC & $51$ atoms EC&PA \\\hline
$(-1, 0, 0)$&-3.39 &-3.36 &-1.86\\
$(0,\pm 1, 0)$& -0.03& -0.03& 0 \\
$(0, 0, \pm1)$& -0.03&-0.03 &0 \\
$(-1, 0, 0)$&3.159 &3.12 &1.86\\
\end{tabular}
\vspace{0.5cm}\\
\begin{tabular}{c|cc|c}
neighbor &$27$ atoms EC & $51$ atoms EC&PA \\\hline
$(-1, 0, 0)$&-0.09 &-0.09 &0\\
$(0,-1, 0)$& -0.28& -0.28& 0.38 \\
$(0, 0, \pm1)$& 0&0 &0 \\
$(0,-1, 0)$& 0.28& 0.28& 0.38 \\
$(-1, 0, 0)$&0.09 &0.09 &0\\
\end{tabular}
\caption{Average absolute value of the off-diagonal SLC parameters $J_{ij,i}^{\mathrm{diag},x}=\frac{1}{2}\left(J_{ij,i}^{xx,x}+J_{ij,i}^{yy,x} \right)$ (top) and $J_{ij,i}^{\mathrm{off},x}=\frac{1}{2}\left(J_{ij,i}^{xy,x}+J_{ij,i}^{yx,x} \right)$ (bottom)  in meV/a.u. for next-nearest neighbors, $i=k$ and a displacement in $x$ direction in bcc Fe obtained by the embedded cluster (EC) method for clusters with $27$ and $51$ atoms and for the closed SLC expressions (averages over the values for the neighbors listed in each line). For the EC method $u_i^x = 0.005\,a_{\mathrm{lat}}$ was used. }
\label{tab:SLC_4}
\end{table}

\section{Fourier Transformed SLC \label{appendix:FT}}

Following a similar procedure as e.g. proposed by Refs. \cite{Rueckriegel2014,Rueckriegel2020, Mankovsky2022} the magnetoelastic anisotropy energy due to DMI will be expressed in terms of spin creation and annihilation operators. Introducing the spin lowering and raising operators into the DMI part of the SLC Hamiltonian Eq. \eqref{eq:Hamilt_extended_magneto-elastic} yields
\begin{align*}
    &H_{\mathrm{SLC-DMI}} =\frac{1}{S^2} \sum_{ij}\sum_{k,\mu} \left[ \vec{D}_{ij,k}^{\mu} \left( \vec{\hat{S}_i}\times \vec{\hat{S}}_j\right)\right]u_k^{\mu} \\
    &\quad=\frac{i}{2S^2}\sum_{ijk, \mu} \big[ D_{ij,k}^{-\mu} \left( \hat{S}_i^z \hat{S}_j^{+}-\hat{S}_i^{+} \hat{S}_j^z\right)\\
    &\quad\quad\quad\quad\quad\quad+D_{ij,k}^{+\mu} \left( \hat{S}_i^{-} \hat{S}_j^z-\hat{S}_i^z \hat{S}_j^{-}\right) \\
    &\quad\quad\quad\quad\quad\quad+D_{ij,k}^{z\mu} \left( \hat{S}_i^+ \hat{S}_j^{-}-\hat{S}_i^{-} \hat{S}_j^+\right)\big]u_k^{\mu} \, ,
\end{align*}
where the orientation vectors of the magnetic moments from the previous sections are replaced by normalized spin operators $\hat{\vec{S}}_{i,j} \to \frac{1}{S}\hat{\vec{S}}_{i,j}$. This can be rewritten in terms of creation and annihilation operators when using a Holstein-Primakoff transformation \cite{Holstein1940} to
\begin{align*}
    H_{\mathrm{SLC-DMI}}
    &=\frac{i}{S}\sum_{ijk,\mu} \Big[ D_{ij,k}^{-\mu} \sqrt{\frac{S}{2}}\left( \hat{b}_j -\hat{b}_i\right)\\
    &\quad\quad\quad\quad+D_{ij,k}^{+\mu}\sqrt{\frac{S}{2}}\left( \hat{b}_i^{\dagger} -\hat{b}_j^{\dagger}\right)\\
    &\quad\quad\quad\quad+D_{ij,k}^{z\mu} \left( \hat{b}_i\hat{b}_j^{\dagger}-\hat{b}_i^{\dagger}\hat{b}_j\right)\Big]u_k^{\mu}. 
\end{align*}
By applying a Fourier transformation for the bosonic creation and annihilation operators
and lattice distortions one can define the Fourier transforms of the DMI components as follows:
\begin{align}
    D_{i,\vec{q}}^{x(y),\mu}=\sum_{\vec{R}_{ji}, \vec{R}_{ki}}D_{ij,k}^{x(y),\mu}e^{i\vec{q}\cdot\vec{R}_{ji}}e^{-i\vec{q}\cdot \vec{R}_{ki}} 
\end{align}
and
\begin{align}
    D_{i,\vec{k}\vec{k}^{\prime}}^{z,\mu}=\sum_{\vec{R}_{ik}, \vec{R}_{jk}}D_{ij,k}^{z,\mu}e^{-i\vec{k}\cdot\vec{R}_{ik}}e^{i\vec{k}^{\prime}\cdot\vec{R}_{jk}}.
    \label{eq:DMI(q)_2}
\end{align}
All together this results in
\begin{align}
    H_{\mathrm{SLC-DMI}}=&\frac{2i}{\sqrt{2S}}\sum_{\mu}\sum_{\vec{q}}\Big[ D_{\vec{q}}^{-\mu}\hat{b}_{\vec{q}} -D_{-\vec{q}}^{+\mu}\hat{b}^{\dagger}_{-\vec{q}}\Big]u_{\vec{q}}^{\mu}\\
    &
    -\frac{2i}{\sqrt{N}S}\sum_{\vec{k},\vec{k}^{\prime}}\sum_{\mu}D_{\vec{k}, \vec{k}^{\prime}}^{z\mu}\hat{b}_{\vec{k}}^{\dagger}\hat{b}_{\vec{k}^{\prime}}u_{(\vec{k}^{\prime}-\vec{k})}^{\mu}+\dots \,.
    \label{eq:H_SLC-DMI(q)}
\end{align}

As discussed by Refs. \cite{Rueckriegel2014, Rueckriegel2020, Streib2019} the first two terms $D_{\vec{q}}^{-\mu}\hat{b}_{\vec{q}}$ ($D_{-\vec{q}}^{+\mu}\hat{b}_{-\vec{q}}^{\dagger}$) describe the interaction of a phonon and magnon, where a magnon is annihilated (created). The last term in Eq. \eqref{eq:H_SLC-DMI(q)}, $D_{\vec{k}, \vec{k}^{\prime}}^{z\mu}\hat{b}_{\vec{k}}^{\dagger}\hat{b}_{\vec{k}^{\prime}}$, represent the magnon-number conserving scattering processes since one magnon is annihilated and one is created \cite{Rueckriegel2020}. Angular momentum can only be transferred between spins and lattice by magnon-number non-conserving processes (i.e. $D_{\vec{q}}^{-\mu}\hat{b}_{\vec{q}}$  and $D_{-\vec{q}}^{+\mu}\hat{b}_{-\vec{q}}^{\dagger}$).



To compare the SLC-DMI to other contributions of the SLC tensor a similar manipulation including the Fourier transformation of spin orientation vectors and displacements yields
\begin{align}
    H_{\mathrm{SLC}} = \frac{1}{\sqrt{N}} \sum_{\vec{k}\vec{k}^{\prime},\vec{q}}\sum_{\alpha \beta, \mu}J_{\vec{k},\vec{k}^{\prime}}^{\alpha \beta, \mu}e_{\vec{k}}^{\alpha}e_{\vec{k}^{\prime}}^{\beta}u_{\vec{q}}^{\mu}\delta_{\vec{k}, \vec{k}^{\prime}+\vec{q}}+\dots \;,
    \label{eq:H_SLC_FT}
\end{align}

with
\begin{align}
   J_{i,\vec{k}\vec{k}^{\prime}}^{\alpha \beta, \mu} = \sum_{\vec{R}_{ik},\vec{R}_{jk}}J_{ij,k}^{\alpha \beta, \mu} e^{-i\vec{k}\cdot \vec{R}_{ik}} e^{i\vec{k}^{\prime}\cdot \vec{R}_{jk}}.
\end{align}

Analogously, we find for the on-site parameters with $i=j$ that $\vec{k}^{\prime}=0$ in Eq. \eqref{eq:H_SLC_FT} and 
\begin{align}
   J_{i,\vec{q}}^{\alpha \beta, \mu} = \sum_{\vec{R}_{ik}}J_{ii,k}^{\alpha \beta, \mu} e^{-i\vec{q}\cdot \vec{R}_{ik}}.
\end{align}
In inversion symmetric crystals all contributions $D_{\vec{q}}^{\pm \mu}$, $D_{\vec{k}, \vec{k}^{\prime}}^{z\mu}$, $J_{\vec{k}, \vec{k}^{\prime}}^{\alpha \beta,\mu}$ and $J_{\vec{q}}^{\alpha \beta,\mu}$ are purely imaginary.

\section{Connection to magnon and phonon dispersions}

\begin{figure}
\centering
   \includegraphics[width=0.49\textwidth]{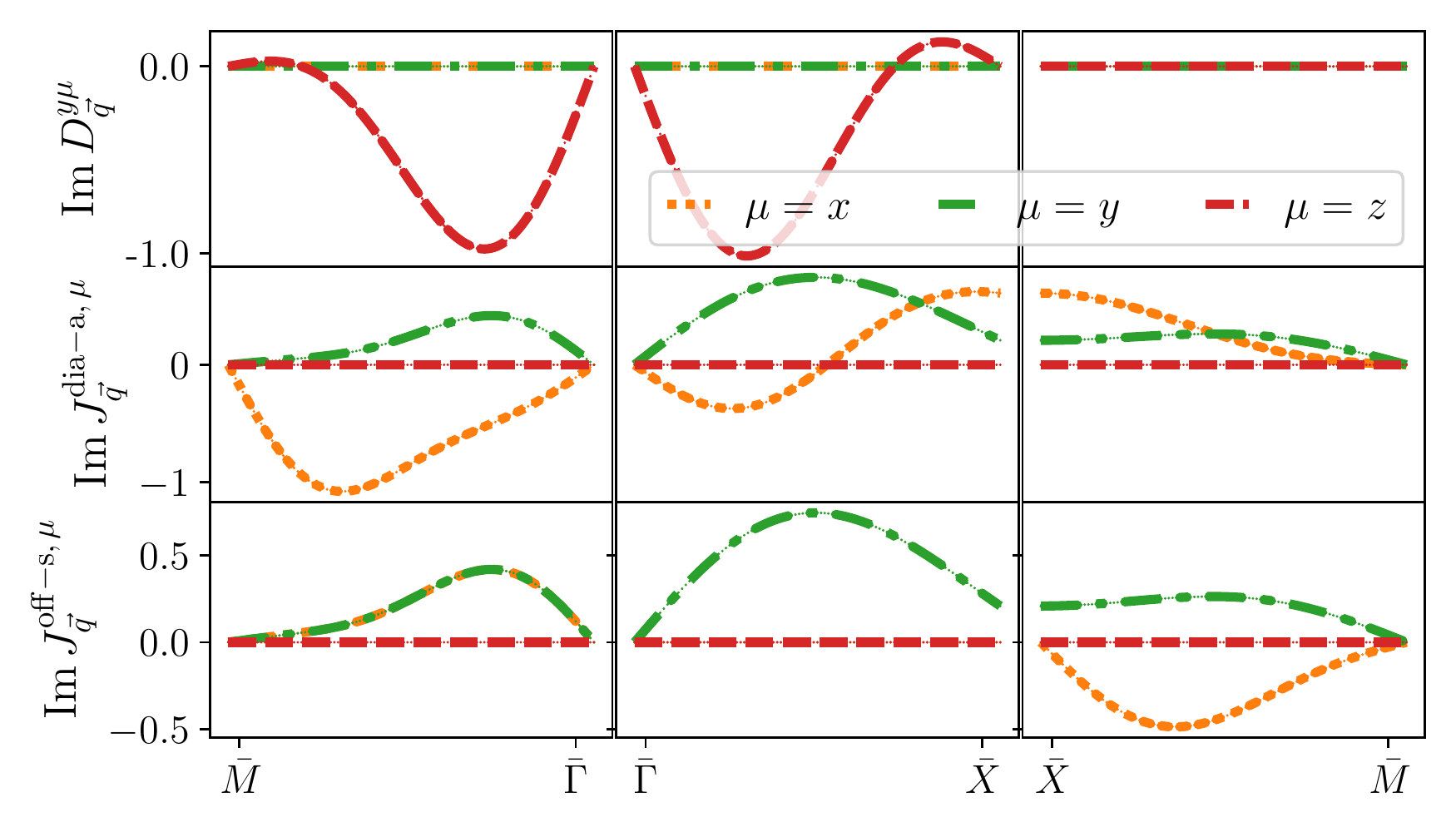}
    \caption[Comparison of Fourier transformed SLC contributions in a Fe(001) monolayer.]{Imaginary part of SLC-DMI $\mathrm{Im}\,D_{\vec{q}}^{x\mu}$ (top), anti-symmetric diagonal elements $\mathrm{Im}\,J_{\vec{q}}^{\mathrm{diag-a},\mu}$ (middle) and symmetric off-diagonal elements $\mathrm{Im}\,J_{\vec{q}}^{\mathrm{off-s},\mu}$ (bottom) of the SLC parameters in meV/a.u. for a Fe(001) monolayer, with $\mu = x, y, z$, plotted for $\vec{q}$ along high-symmetry lines of the Brillouin zone.
    The real part is zero in all cases. Neighbors with $r_{ij}\leq 3.0\,a_{\mathrm{lat}}$ are considered.}
\label{fig:FT_SLC_contributions_Fe001}
\end{figure}

To investigate the impact of spin-lattice interactions on magnon and phonon modes the Fourier transformed spin-lattice contributions can be calculated (for further details see Ref. \cite{Mankovsky2022}). As an example, the results for Fe(001) are shown in Fig. \ref{fig:FT_SLC_contributions_Fe001}. In the top panel the Fourier transformed SLC-DMI 
\begin{align}
    D_{i,\vec{q}}^{x(y),\mu}=\sum_{\vec{R}_{ji}, \vec{R}_{ki}}D_{ij,k}^{x(y),\mu}e^{i\vec{q}\cdot\vec{R}_{ji}}e^{-i\vec{q}\cdot \vec{R}_{ki}},
    \label{eq:DMI(q)_1}
\end{align}
in the middle and bottom panels the Fourier transformed SLC related to on-site anisotropy
\begin{align}
   J_{i,\vec{q}}^{\alpha \beta, \mu} = \sum_{\vec{R}_{ik}}J_{ii,k}^{\alpha \beta, \mu} e^{-i\vec{q}\cdot \vec{R}_{ik}}.
\end{align}
are presented. These quantities represent the modification of magnon and phonon modes for a hybridization of these modes. For further details see Appendix \ref{appendix:FT} and Ref. \cite{Mankovsky2020}. Note that in contrast to the SLC presented in Tab. \ref{tab:Fe_bulk_monolayers_SLC} the Fourier transformations include the sum over neighboring atoms up to a distance of $3.0\,a_{\mathrm{lat}}$ to atom $i$ and different configurations of the displaced atom. As discussed by Refs. \cite{Rueckriegel2014, Rueckriegel2020, Streib2019, Mankovsky2020} the SLC-DMI $D_{\vec{q}}^{x(y)\mu}$ describe the interaction strength of a phonon and magnon, where a magnon is annihilated (created) and hence contribute to an angular momentum transfer. This is explained in further detail in Appendix \ref{appendix:FT}.

\section{Splitting of the SLC according to exchange mechanism \label{sec:spit-SLC}}

Representing the isotropic exchange coupling parameters explicitely
in terms of two contributions due to direct exchange (de) and
superexchange (se), $J_{ij} =
J^{\mathrm{de}}_{ij} + J^{\mathrm{se}}_{ij}$, one can see their impact
also to different SLC parameters, assuming that $J^{\mathrm{de}}_{ij}$
depends on the distance $|\vec{R}_{ij}|$ and $J^{\mathrm{se}}_{ij}$
depends on the angle $\alpha(Cr_1-O(S)-Cr_2)$. Variarion of the distance by
$u = \delta |\vec{R}_{ij}|$  results in the change of the direct contribution
\begin{align}
  J^{\mathrm{de}}_{ij} ( u) = J^{\mathrm{de},0}_{ij} + \left. \frac{\partial}{\partial u} J^{\mathrm{de}}_{ij}(u)\right\vert_{u=0} u \, ,  
  \label{eq:Mod_J_simplified_picture1}
\end{align}
as well as the superexchange contribution (taking into account that the changes of the angle $\alpha$ due to displacement of Cr atom)
\begin{align}
    J^{\mathrm{se}}_{ij}(\alpha +
  \Delta\alpha) =  J^{\mathrm{se},0}_{ij} + \left.\frac{\partial
  J^{\mathrm{se}}_{ij}(\alpha)}{\partial \alpha}\right \vert_{u=0} \frac{\partial
  \alpha}{\partial u} u\, ,  
  \label{eq:Mod_J_simplified_picture2}
\end{align}
In the case of $\vec{R}_{ij} = (0,a,0)$, the derivative 
$\frac{\partial}{\partial u} J^{\mathrm{de}}_{ij}(u)$ is defined by the
SLC parameter  ${\cal J}^{\mathrm{se},y}_{ij,j}$.

On the other hand, varying the distance $|\vec{R}_{ik}|$ with $k \neq i(j)$,
e.g. along $z$ direction, only the superexchange contribution gets a contribution linear w.r.t. the displacement $v^z_k = \delta R^z_{ik}$
\begin{align}
   J^{\mathrm{se}}_{ij}(\alpha +
  \Delta\alpha) =  J^{\mathrm{se}}_{ij} + \left.\frac{\partial
  J^{\mathrm{se}}_{ij}(\alpha)}{\partial \alpha} \right \vert_{u=0} \frac{\partial
  \alpha}{\partial v^z_k} v^z_k\, ,  
  \label{eq:Mod_J_simplified_picture3}
\end{align}
In this case the derivative 
$\frac{\partial}{\partial v^z_k} J^{\mathrm{se}}_{ij}$ is defined by the
SLC parameter  ${\cal J}^{\mathrm{se},z}_{ij,k}$.




\end{document}